\documentclass{tlp}
\usepackage{amsmath}
\usepackage{amssymb}
\usepackage{stmaryrd}
\usepackage{epsfig}
\usepackage{floatflt}
\usepackage{proof}
\title[Mapping Fusion and SHR into Logic Programming]{Mapping Fusion and 
Synchronized Hyperedge Replacement into Logic Programming\thanks{Work 
supported in part by the European IST-FET Global Computing project IST-2001-33100
PROFUNDIS and the European IST-FET Global
Computing 2 project IST-2005-16004 {\sc Sensoria}.}}
\submitted{27 December 2003}
\revised{14 April 2005}
\accepted{5 January 2006}

\author[Ivan Lanese and Ugo Montanari]{IVAN LANESE and UGO 
MONTANARI\\Dipartimento di Informatica,
Universit\`a di
Pisa\\Largo Bruno Pontecorvo,3 56127 Pisa, Italia\\\email{\{lanese,ugo\}@di.unipi.it}}

\newcommand{\ie}{i.e.\ }
\newcommand{\wrt}{w.r.t.\ }
\newcommand{\eg}{e.g.\ }

\newtheorem{defin}{Definition}
\newtheorem{theorem}{Theorem}
\newtheorem{ex}{Example}

\newtheorem{lemma}{Lemma}

\newcommand{\transl}[1]{\llbracket #1 \rrbracket}

\newcommand{\xRightarrow}[1]{\stackrel{#1}{\Rightarrow}}
\DeclareMathOperator{\grammar}{::=}
\DeclareMathOperator{\fn}{fn}
\DeclareMathOperator{\bn}{bn}
\DeclareMathOperator{\n}{n}

\DeclareMathOperator{\fnarray}{fnarray}
\DeclareMathOperator{\im}{Im}
\DeclareMathOperator{\dom}{dom}
\DeclareMathOperator{\act}{act}
\DeclareMathOperator{\ar}{ar}

\DeclareMathOperator{\rank}{rank}
\DeclareMathOperator{\mgu}{mgu}

\DeclareMathOperator{\eqn}{eqn}
\DeclareMathOperator{\rec}{rec}

\DeclareMathOperator{\bigpar}{\big[\!\!\big]}
\newcommand{\eqnum}[1]{\stackrel{#1}{=}}
\DeclareMathOperator{\mstar}{\!\!\mbox{}^*\,\,}

\begin{document}
\maketitle
\begin{abstract}
In this paper we compare three different formalisms that can be used
in the area of models for distributed, concurrent and mobile
systems. In particular we analyze the relationships between a process
calculus, the \emph{Fusion Calculus}, graph transformations in the
\emph{Synchronized Hyperedge Replacement} with Hoare synchronization
(HSHR) approach and \emph{logic programming}. We present a translation
from Fusion Calculus into HSHR (whereas Fusion Calculus uses Milner
synchronization) and prove a correspondence between the reduction
semantics of Fusion Calculus and HSHR transitions. We also present a
mapping from HSHR into a transactional version of logic programming
and prove that there is a full correspondence between the two
formalisms. The resulting mapping from Fusion Calculus to logic
programming is interesting since it shows the tight analogies between
the two formalisms, in particular for handling name generation and
mobility. The intermediate step in terms of HSHR is convenient since
graph transformations allow for multiple, remote synchronizations, as
required by Fusion Calculus semantics. To appear in Theory and
Practice of Logic Programming (TPLP).
\end{abstract}


\section{Introduction}
In this paper we compare different formalisms that can be used to specify 
and model systems which are distributed, concurrent and mobile, as those 
that are usually found in the global computing area.\\
Global computing is becoming very important because of the great 
development of networks which are deployed on huge areas, first of all 
Internet, but also other kinds of networks such as networks for wireless 
communications. In order to build and program these networks one needs to 
deal with issues such as reconfigurability, synchronization and 
transactions at a suitable level of abstraction. Thus powerful formal 
models and tools are needed. Until now no model has been able to emerge as 
the standard one for this kind of systems, but there are a lot of 
approaches with different merits and drawbacks.\\

An important approach is based on process calculi, like Milner's CCS and 
Hoare's CSP. These two calculi deal with communication and synchronization 
in a simple way, but they lack the concept of mobility. An important 
successor of CCS, the $\pi$-calculus \cite{milnermobile}, allows to study 
a wide range of mobility problems in a simple mathematical framework. 
We are mainly interested in the \emph{Fusion Calculus} 
\cite{victorfusion,tesivictor,gardnerexplicit,gardnerbisimulation}, which 
is an evolution of $\pi$-calculus. The interesting aspect of this calculus 
is that it has been obtained by simplifying and making more symmetric the 
$\pi$-calculus.\\

One of the known limitations of process-calculi when applied to
distributed systems is that they lack an intuitive representation
because they are equipped with an \emph{interleaving} semantics and
they use the same constructions for representing both the agents and
their configurations. An approach that solves this kind of problems is
based on \emph{graph transformations} \cite{graphbook3}. In this case
the structure of the system is explicitly represented by a graph which
offers both a clean mathematical semantics and a suggestive
representation. In particular we represent computational entities such
as processes or hosts with hyperedges (namely edges attached to any
number of nodes) and channels between them with shared nodes. As far
as the dynamic aspect is concerned, we use \emph{Synchronized
Hyperedge Replacement} with \emph{Hoare synchronization} (HSHR)
\cite{deganomodel}. This approach uses productions to specify the
behaviour of single hyperedges, which are synchronized by exposing
actions on nodes. Actions exposed by different hyperedges on the same
node must be compatible. In the case of Hoare synchronization all the
edges must expose the same action (in the CSP style). This approach
has the advantage, \wrt other graphical frameworks such as Double
Pushout \cite{ehrigDPO} or Bigraphs \cite{milnerbigraphs}, of allowing a
distributed implementation since productions have a local effect and
synchronization can be performed using a distributed algorithm.\\ 
We use the extension of HSHR with \emph{mobility}
\cite{hirschreconfiguration,hirschsynchronized,konigobservational,tuostoambients,miatesi},
that allows edges to expose node references together with actions, and
nodes whose references are matched during synchronization are
unified.\\

For us HSHR is a good step in the direction of \emph{logic programming} \cite{lloydLP}. 
We consider logic programming as a formalism for modelling concurrent and 
distributed systems. This is a non-standard view of logic programming (see 
\citeNP{bruniinteractive} for a presentation of our approach) which 
considers goals as processes whose evolution is defined by Horn clauses 
and whose interactions use variables as channels and are managed by the 
unification engine. In this framework we are not interested only in 
refutations, but in any partial computation that rewrites a goal into 
another.\\

In this paper we analyze the relationships between these three formalisms 
and we find tight analogies among them, like the same parallel composition 
operator and the use of unification for name mobility. However we also 
emphasize the differences between these models:
\begin{itemize}
\item the Fusion Calculus is interleaving and relies on \emph{Milner 
synchronization} (in the CCS style);
\item HSHR is inherently \emph{concurrent} since many actions can
be performed at the same time on different nodes and uses Hoare
synchronization;
\item logic programming is concurrent, has a wide spectrum of possible 
controls which are based on the Hoare synchronization model, and also is 
equipped with a more complex data management.
\end{itemize}
We will show a mapping from Fusion Calculus to HSHR and prove a
correspondence theorem. Note that HSHR is a good intermediate step
between Fusion Calculus and logic programming since in HSHR hyperedges
can perform multiple actions at each step, and this allows to build
chains of synchronizations. This additional power is needed to model
Milner synchronization, which requires synchronous, atomic routing
capabilities. To simplify our treatment we consider only reduction
semantics. The interleaving behaviour is imposed with an external
condition on the allowed HSHR transitions.\\

Finally we present the connections between HSHR and logic programming.
Since the logic programming paradigm allows for many computational
strategies and is equipped with powerful data structures, we need to
constrain it in order to have a close correspondence with HSHR. We
define to this end \emph{Synchronized Logic Programming} (SLP), which
is a transactional version of logic programming. The idea is that
function symbols are pending constraints that must be satisfied before
a transaction can commit, as for zero tokens in zero-safe nets
\cite{zero-safe}. In the mapping from HSHR to SLP edges are translated
into predicates, nodes into variables and parallel composition into
AND composition.\\

This translation was already presented in the MSc.\ thesis of the
first author \cite{miatesi} and in \citeN{lanesesoftware}. Fusion
Calculus was mapped into SHR with Milner synchronization (a simpler
task) in \citeN{ivancometa} where Fusion LTS was considered instead of
Fusion reduction semantics. The paper \citeN{lanesesoftware} also
contains a mapping of Ambient calculus into HSHR. This result can be
combined with the one here, thus obtaining a mapping of Ambient
calculus into SLP. An extensive treatment of all the topics in this
paper can also be found in the forthcoming Ph.D. thesis of the first
author \cite{miaphdtesi}.\\
 
Since logic programming is not only a theoretical framework, but also
a well developed programming style, the connections between Fusion,
HSHR and logic programming can be used for implementation
purposes. SLP has been implemented in \citeN{miatesi} through
meta-interpretation. Thus we can use translations from Fusion and HSHR
to implement them. In particular, since implementations of logic
programming are not distributed, this can be useful mainly for
simulation purposes.\\

In Section \ref{section:background} we present the required background, in 
particular we introduce the Fusion Calculus (\ref{subsection:fusion}), the 
algebraic representation of graphs and the HSHR (\ref{subsection:SHR}), 
and logic programming (\ref{subsection:logicprog}). Section 
\ref{section:fusion2SHR} is dedicated to the mapping from Fusion Calculus 
to HSHR. Section \ref{section:HSHR2lp} analyzes the relationships between 
HSHR and logic programming, in particular we introduce SLP 
(\ref{subsection:SLP}), we prove the correspondence between it and HSHR 
(\ref{subsection:HSHR2lp}) and we give some hints on how to implement 
Fusion Calculus and HSHR using Prolog (\ref{subsection:meta}). In 
Section \ref{section:conclusion} we present some conclusions and traces 
for future work. Finally, proofs and technical lemmas are in \ref{appendix:proofs}.
\section{Background}\label{section:background}
\paragraph{Mathematical notation.}
{\small We use $T\sigma$ to denote the application of substitution
$\sigma$ to $T$ (where $T$ can be a term or a set/vector of terms).
We write substitutions as sets of pairs of the form $t/x$, denoting
that variable $x$ is replaced by term $t$.  We also denote with
$\sigma_1\sigma_2$ the composition of substitutions $\sigma_1$ and
$\sigma_2$. We denote with $\sigma^{-1}(x)$ the set of elements mapped
to $x$ by $\sigma$. We use $|-|$ to denote the operation that computes
the number of elements in a set/vector. Given a function $f$ we denote
with $\dom(f)$ its domain, with $\im(f)$ its image and with $f|_S$ the
restriction of $f$ to the new domain $S$. We use on functions and
substitutions set theoretic operations (such as $\cup$) referring to
their representation as sets of pairs. Similarly, we apply them to
vectors, referring to the set of the elements in the vector. In
particular, $\setminus$ is set difference. Given a set $S$ we denote
with $S^*$ the set of strings on $S$. Also, given a vector $\vec v$
and an integer $i$, $\vec v[i]$ is the $i$-th element of $\vec
v$. Finally, a vector is given by listing its elements inside angle
brackets $\langle - \rangle$.}
\subsection{The Fusion Calculus}\label{subsection:fusion}
The Fusion Calculus \cite{victorfusion,tesivictor} is a calculus for 
modelling distributed and mobile systems which is based on the concepts of 
\emph{fusion} and scope. It is an evolution of the $\pi$-calculus 
\cite{milnermobile} and the interesting point is that it is obtained by 
simplifying the calculus. In fact the two action prefixes for input and 
output communication are symmetric, whereas in the $\pi$-calculus they are 
not, and there is just one binding operator called scope, whereas the 
$\pi$-calculus has two (restriction and input). As shown in 
\citeNS{victorfusion}, the $\pi$-calculus is syntactically a subcalculus 
of the Fusion Calculus (the key point is that the input of $\pi$-calculus 
is obtained using input and scope). In order to have these properties 
fusion actions have to be introduced. An asynchronous version of Fusion 
Calculus is described in \citeN{gardnerexplicit}, 
\citeN{gardnerbisimulation}, where name fusions are handled explicitly as 
messages. Here we follow the approach by Parrow and Victor.\\

We now present in details the syntax and the reduction semantics of
Fusion Calculus. In our work we deal with a subcalculus of the Fusion
Calculus, which has no match and no mismatch operators, and has only
guarded summation and recursion. All these restrictions are quite
standard, apart from the one concerning the match operator, which is
needed to have an expansion lemma. To extend our approach to deal with
match we would need to extend SHR by allowing production applications
to be tagged with a unique identifier. We leave this extension for
future work. In our discussion we distinguish between \emph{sequential
processes} (which have a guarded summation as topmost operator) and
general processes.\\

We assume to have an infinite set $\mathcal{N}$ of names ranged over
by $u,v,\dots,z$ and an infinite set of agent variables (disjoint
\wrt the set of names) with meta-variable $X$. Names represent
communication channels. We use $\phi$ to denote an equivalence
relation on $\mathcal{N}$, called fusion, which is represented in the
syntax by a finite set of equalities. Function $\n(\phi)$ returns all
names which are fused, \ie those contained in an equivalence class
of $\phi$ which is not a singleton.

\begin{defin}
The prefixes are defined by:
\begin{eqnarray*}
\alpha & \grammar & u \vec x \qquad \textrm{(Input)}\\
& & \overline{u} \vec x \qquad \textrm{(Output)}\\
& & \phi \qquad \ \ \textrm{(Fusion)}
\end{eqnarray*}
\end{defin}

\begin{defin}
The agents are defined by:
\begin{eqnarray*}
S \grammar & \sum_i \alpha_i.P_i & \textrm{(Guarded sum)}
\end{eqnarray*}
\begin{eqnarray*}
P \grammar & 0 & \textrm{(Inaction)}\\
& S & \textrm{(Sequential Agent)}\\
& P_1|P_2 & \textrm{(Composition)}\\
& (x) P & \textrm{(Scope)}\\
& \rec X.P & \textrm{(Recursion)}\\
& X & \textrm{(Agent variable)}
\end{eqnarray*}
\end{defin}

The scope restriction operator is a binder for names, thus $x$ is
bound in $(x)P$.  Similarly $\rec$ is a binder for agent variables. We
will only consider agents which are closed \wrt both names and agent
variables and where in $\rec X.P$ each occurrence of $X$ in $P$ is
within a sequential agent (guarded recursion). We use recursion to
define infinite processes instead of other operators (\eg replication)
since it simplifies the mapping and since their expressive power is
essentially the same. We use infix $+$ for binary sum (which thus is
associative and commutative).\\

Given an agent $P$, functions $\fn$, $\bn$ and $\n$ compute the sets 
$\fn(P)$, $\bn(P)$ and $\n(P)$ of its free, bound and all names 
respectively.\\

Processes are agents considered up to structural axioms defined as follows.
\begin{defin}[Structural congruence]\label{defin:congrfusion}
The structural congruence $\equiv$ between agents is the least congruence 
satisfying the $\alpha$-conversion law (both for names and for agent 
variables), the abelian monoid laws for composition 
(associativity, commutativity and 0 as identity), the scope laws $(x)0 
\equiv 0$, $(x)(y)P \equiv (y)(x)P$, the scope extrusion law $P|(z)Q 
\equiv (z)(P|Q)$ where $z \notin \fn(P)$ and the recursion law $\rec X.P 
\equiv P\{\rec X.P/X\}$. 
\end{defin}

Note that $\fn$ is also well-defined on processes.

In order to deal with fusions we need the following definition.
\begin{defin}[Substitutive effect]
A substitutive effect of a fusion $\phi$ is any idempotent substitution 
$\sigma:\mathcal{N} \rightarrow \mathcal{N}$ having $\phi$ as its kernel. 
In other words $x\sigma=y\sigma$ iff $x \phi y$ and $\sigma$ sends all 
members of each equivalence class of $\phi$ to one representative in the 
class\footnote{Essentially $\sigma$ is a most general unifier of $\phi$, 
when it is considered as a set of equations.}.
\end{defin}

The reduction semantics for Fusion Calculus is the least relation 
satisfying the following rules.

\begin{defin}[Reduction semantics for Fusion 
Calculus]\label{defin:redfusion}
$$(\vec z) (R|(\dots + u \vec x .P)|(\overline{u} \vec y.Q+\dots)) 
\rightarrow (\vec z)(R|P|Q)\sigma$$
where $|\vec x|=|\vec y|$ and $\sigma$ is a substitutive effect of $\{ 
\vec x = \vec y \}$ such that $\dom(\sigma) \subseteq \vec z$.
$$(\vec z) (R|(\dots + \phi .P)) \rightarrow (\vec z)(R|P)\sigma$$
where $\sigma$ is a substitutive effect of $\phi$ such that $\dom(\sigma) \subseteq \vec z$.
$$\frac{P \equiv P', P' \rightarrow Q' , Q' \equiv Q}{P \rightarrow Q}$$
\end{defin}

\subsection{Synchronized Hyperedge Replacement}\label{subsection:SHR}
Synchronized Hyperedge Replacement (SHR) \cite{deganomodel} is an approach 
to (hyper)graph transformations that defines \emph{global transitions} 
using \emph{local productions}. Productions define how a single (hyper)edge 
can be rewritten and the \emph{conditions} that this rewriting imposes on 
adjacent nodes. Thus the global transition is obtained by applying in 
parallel different productions whose conditions are compatible. What 
exactly compatible means depends on which \emph{synchronization model} we 
use. In this work we will use the Hoare synchronization model (HSHR), 
which requires that all the edges connected to a node expose the same 
action on it. For a general definition of synchronization models see \citeN{ivanFGUC}.\\
We use the extension of HSHR with \emph{mobility} 
\cite{hirschreconfiguration,hirschsynchronized,konigobservational,tuostoambients,miatesi}, 
that allows edges to expose node references together with actions, and 
nodes whose references are matched during synchronization are unified.\\

We will give a formal description of HSHR as labelled
transition system, but first of all we need an algebraic
representation for graphs.

An edge is an atomic item with a label and with as many ordered
tentacles as the rank $\rank(L)$ of its label $L$. A set of nodes, together with a
set of such edges, forms a graph if each edge
is connected, by its tentacles, to its attachment nodes. We will
consider graphs up to isomorphisms that preserve\footnote{In our
approach nodes usually represent free names, and they are preserved by
isomorphisms.}  nodes, labels of edges, and connections between edges
and nodes.\\ 
Now, we present a definition of graphs as syntactic
judgements, where nodes correspond to names and edges to basic terms
of the form $L(x_1,\dots,x_n)$, where the $x_i$ are arbitrary names
and $\rank(L)=n$. Also, $nil$ represents the empty graph and
$|$ is the parallel composition of graphs (merging nodes with the same
name).

\begin{defin}[Graphs as syntactic judgements]
Let $\mathcal{N}$ be a fixed infinite set of names and $LE$ a ranked 
alphabet of labels. A syntactic judgement (or simply a judgement) is of 
the form $\Gamma \vdash G$ where:
\begin{enumerate}
\item $\Gamma \subseteq \mathcal{N}$ is the (finite) set of nodes in the 
graph.
\item $G$ is a term generated by the grammar\\
$G \grammar L(\vec x) \ | \ G_1|G_2 \ | \ nil$\\ 
where $\vec x$ is a vector of names and $L$ is an edge label with 
$\rank(L)=|\vec x|$.
\end{enumerate}

\end{defin}
We denote with $\n$ the function that given a graph $G$ returns the set 
$\n(G)$ of all the names in $G$.
We use the notation $\Gamma,x$ to denote the set obtained by adding $x$ to 
$\Gamma$, assuming $x \notin \Gamma$. Similarly, we write 
$\Gamma_1,\Gamma_2$ to state that the resulting set of names is the 
disjoint union of $\Gamma_1$ and $\Gamma_2$.

\begin{defin}[Structural congruence and well-formed judgements] 
\label{defin:structural}

The structural congruence $\equiv$ on terms $G$ obeys the following axioms:
$$\textrm{(AG1)} \ (G_1|G_2)|G_3 \equiv G_1|(G_2|G_3)$$  
$$\textrm{(AG2)} \ G_1|G_2 \equiv G_2|G_1$$
$$\textrm{(AG3)} \ G|nil \equiv G$$

The well-formed judgements $\Gamma \vdash G$ over $LE$ and 
$\mathcal{N}$ are those where $\n(G) \subseteq \Gamma$.
\end{defin}

Axioms (AG1),(AG2) and (AG3) define respectively the associativity, 
commutativity and identity over $nil$ for operation $|$.\\ 

Well-formed judgements up to structural axioms are isomorphic to graphs up 
to isomorphisms. For a formal statement of the correspondence see \citeN{tesihirsch}.

We will now present the steps of a SHR computation.

\begin{defin}[SHR transition]\label{defin:SHRtrans}
Let $Act$ be a set of actions. For each action $a \in Act$, let  
$\ar(a)$ be its arity.\\
A SHR transition is of the form:
$$\Gamma \vdash G \xrightarrow[]{\Lambda,\pi} \Phi \vdash G'$$
where $\Gamma \vdash G$ and $\Phi \vdash G'$ are well-formed judgements 
for graphs, $\Lambda:\Gamma \rightarrow (Act \times \mathcal{N}^*)$ is a 
total function and $\pi:\Gamma \rightarrow \Gamma$ is an idempotent 
substitution. Function $\Lambda$ assigns to each node $x$ the action $a$ 
and the vector $\vec y$ of node references exposed on $x$ by the 
transition. If $\Lambda(x)=(a,\vec y)$ then we define 
$\act_{\Lambda}(x)=a$ and $\n_{\Lambda}(x)=\vec y$. We require that 
$\ar(\act_{\Lambda}(x))=|\n_{\Lambda}(x)|$, namely the arity of the action 
must equal the length of the vector.\\
We define:
\begin{itemize}
\item $\n(\Lambda)=\{z| \exists x. z \in \n_\Lambda(x)\}$\\ 
set of exposed names;
\item $\Gamma_\Lambda = \n(\Lambda) \setminus \Gamma$\\
set of fresh names that are exposed;
\item $\n(\pi)=\{x| \exists x' \neq x. x\pi=x'\pi\}$\\ 
set of fused names.
\end{itemize}
Substitution $\pi$ allows to merge nodes. Since $\pi$ is idempotent, it 
maps every node into a standard representative of its equivalence class. 
We require that $\forall x \in \n(\Lambda). x\pi=x$, \ie only 
references to representatives can be exposed. Furthermore we require $\Phi 
\supseteq \Gamma\pi \cup \Gamma_{\Lambda}$, namely nodes are never 
erased. Nodes in $\Gamma_{Int}=\Phi \setminus (\Gamma\pi \cup 
\Gamma_{\Lambda})$ are fresh internal nodes, silently created in the 
transition. We require that no isolate, internal nodes are created, 
namely $\Gamma_{Int} \subseteq \n(G')$.
\end{defin}

Note that the set of names $\Phi$ of the resulting graph is fully
determined by $\Gamma$, $\Lambda$, $\pi$ and $G'$ thus we will have no
need to write its definition explicitly in the inference rules. Notice
also that we can write a SHR transition as:
$$\Gamma \vdash G \xrightarrow[]{\Lambda,\pi} 
\Gamma\pi,\Gamma_{\Lambda},\Gamma_{Int} \vdash G'.$$

We usually assume to have an action $\epsilon \in Act$ of arity $0$ to 
denote ``no synchronization''. We may not write explicitly $\pi$ if it is 
the identity, and some actions if they are $(\epsilon,\langle \rangle)$. 
Furthermore we use $\Lambda_{\epsilon}$ to denote the function 
that assigns $(\epsilon,\langle \rangle)$ to each node in $\Gamma$ (note 
that the dependence on $\Gamma$ is implicit). 

We derive SHR transitions from basic productions using a set of inference 
rules. Productions define the behaviour of single edges.

\begin{defin}[Production]\label{defin:production}
A production is a SHR transition of the form:
$$x_1,\dots,x_n \vdash L(x_1,\dots,x_n) \xrightarrow[]{\Lambda,\pi} \Phi 
\vdash G$$
where all $x_i$, $i=1 \dots n$ are distinct.\\ 
Productions are considered as schemas and so they are 
$\alpha$-convertible \wrt names in $\{x_1,\dots,x_n\} \cup 
\Phi$. 
\end{defin}

We will now present the set of inference rules for Hoare synchronization. 
The intuitive idea of Hoare synchronization is that all the edges 
connected to a node must expose the same action on that node. 

\begin{defin}[Rules for Hoare synchronization] \label{defin:hoarerules}

$$\textrm{(par)}\quad \frac{\Gamma \vdash G_1 \xrightarrow[]{\Lambda,
\pi} \Phi \vdash G_2 \quad \Gamma' \vdash G'_1
\xrightarrow[]{\Lambda',\pi'} \Phi' \vdash G'_2 \quad (\Gamma \cup
\Phi) \cap (\Gamma' \cup \Phi')=\emptyset}{\Gamma, \Gamma' \vdash
G_1|G'_1 \xrightarrow[]{\Lambda \cup \Lambda', \pi \cup \pi'} \Phi,
\Phi' \vdash G_2|G'_2}$$

$$\textrm{(merge)}\quad \frac{\Gamma \vdash G_1 \xrightarrow[]{\Lambda, 
\pi} \Phi \vdash G_2 \quad \forall x,y \in \Gamma. x\sigma = y\sigma \wedge x \neq y \Rightarrow \act_{\Lambda}(x) = \act_{\Lambda}(y)}{\Gamma\sigma \vdash G_1\sigma 
\xrightarrow[]{\Lambda', \pi'} \Phi' \vdash G_2\sigma\rho}$$
where $\sigma:\Gamma \rightarrow \Gamma$ is an idempotent substitution and:
\begin{enumerate}[(iii).]
\renewcommand{\theenumi}{(\roman{enumi})}
\item\label{merge:rho} $\rho = \mgu(\{(\n_{\Lambda}(x))\sigma=
(\n_{\Lambda}(y))\sigma|x\sigma= y\sigma \} \cup \{ x\sigma = y\sigma
| x\pi=y\pi\})$ where $\rho$ maps names to representatives in $\Gamma \sigma$ whenever possible
\item\label{merge:lambda} $\forall z \in \Gamma. \Lambda'(z\sigma) = (\Lambda(z))\sigma\rho$
\item\label{merge:pi} $\pi' = \rho|_{\Gamma\sigma}$
\end{enumerate}

$$\textrm{(idle)} \quad \Gamma \vdash G 
\xrightarrow[]{\Lambda_{\epsilon},id} \Gamma \vdash G$$

$$\textrm{(new)} \quad \frac{\Gamma \vdash G_1 \xrightarrow[]{\Lambda,\pi} 
\Phi \vdash G_2 \quad x \notin \Gamma \quad \vec y \cap (\Gamma \cup \Phi \cup \{x\})= \emptyset}{\Gamma,x \vdash G_1 \xrightarrow[]{\Lambda \cup 
\{(x,a,\vec y)\},\pi} \Phi' \vdash G_2}$$
\end{defin}

A transition is obtained by composing productions, which are first
applied on disconnected edges, and then by connecting the edges by
merging nodes. In particular rule (par) deals with the composition of
transitions which have disjoint sets of nodes and rule (merge) allows
to merge nodes (note that $\sigma$ is a projection into
representatives of equivalence classes). The side condition requires
that we have the same action on merged nodes. Definition
\ref{merge:rho} introduces the most general unifier $\rho$ of the
union of two sets of equations: the first set identifies (the
representatives of) the tuples associated to nodes merged by $\sigma$,
while the second set of equations is just the kernel of $\pi$. Thus
$\rho$ is the merge resulting from both $\pi$ and $\sigma$. Note that
\ref{merge:lambda} $\Lambda$ is updated with these merges and that
\ref{merge:pi} $\pi'$ is $\rho$ restricted to the nodes of the graph
which is the source of the transition. Rule (idle) guarantees that
each edge can always make an explicit idle step. Rule (new) allows
adding to the source graph an isolated node where arbitrary actions
(with fresh names) are exposed.\\

We write $\mathcal{P} \Vdash (\Gamma \vdash G \xrightarrow[]{\Lambda,\pi} 
\Phi \vdash G')$ if $\Gamma \vdash G \xrightarrow[]{\Lambda,\pi} \Phi 
\vdash G'$ can be obtained from the productions in $\mathcal{P}$ using 
Hoare inference rules.\\

We will now present an example of HSHR computation.
\begin{ex}[\citeNP{hirschreconfiguration}]
We show now how to use HSHR to derive a 4 elements ring starting from 
a one element ring, and how we can then specify a reconfiguration that 
transforms the ring into the star graph in Figure 
\ref{figure:star4}.

\begin{figure}
\caption{Star graph} \label{figure:star4}
\resizebox{3cm}{!}{\includegraphics{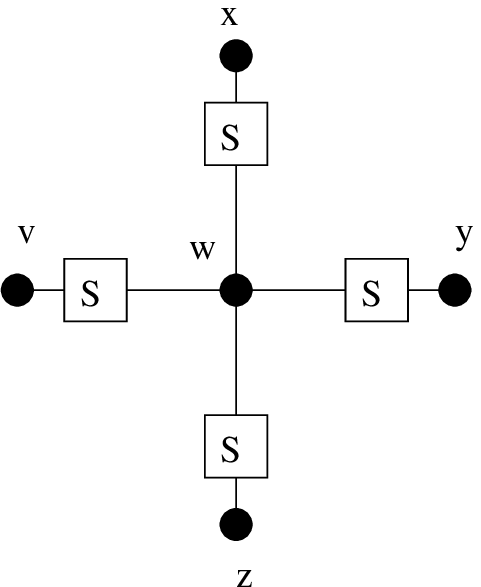}}
\end{figure}

We use 
the following productions:
$$x,y \vdash C(x,y) \xrightarrow[]{(x, \epsilon, \langle \rangle), 
(y, \epsilon, \langle \rangle)} x,y,z \vdash C(x,z)|C(z,y)$$
$$x,y \vdash C(x,y) \xrightarrow[]{(x, r, \langle w \rangle), (y, r, 
\langle w \rangle)} x,y,w \vdash S(y,w)$$
that are graphically represented in Figure \ref{figure:prod}. Notice 
that $\Lambda$ is represented by decorating every node $x$ in the 
left hand with $\act_\Lambda(x)$ and $\n_\Lambda(x)$.
\begin{figure}[!tb] \caption{Productions} \label{figure:prod}
\begin{center}
\resizebox{9cm}{!}{\includegraphics{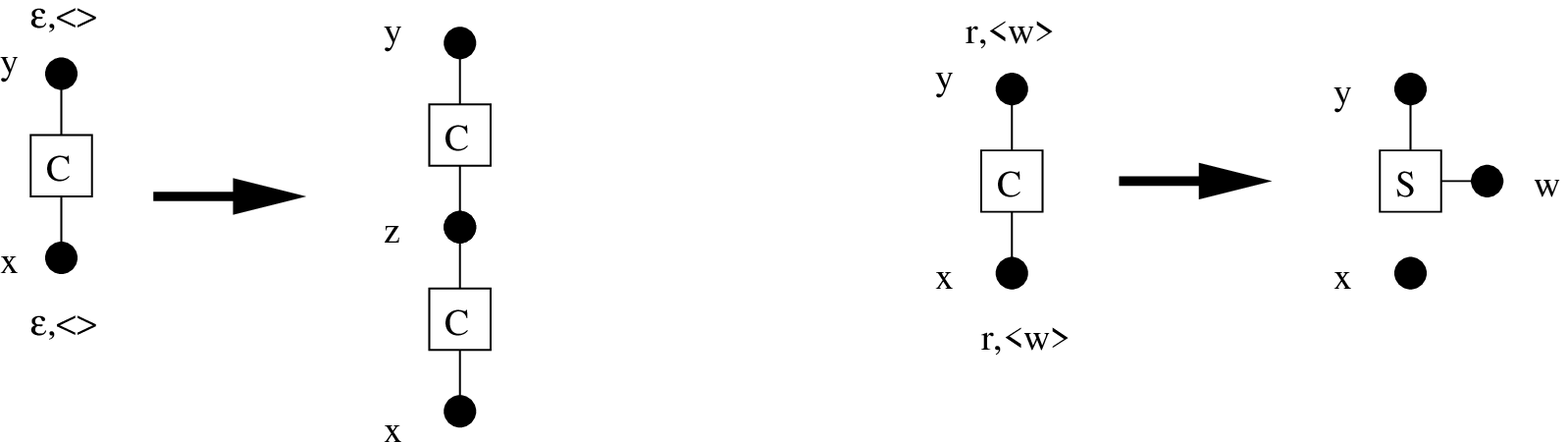}}
\end{center}
\end{figure}
The first rule allows to create rings, in fact we can create all rings 
with computations like:
\begin{multline*}
x \vdash C(x,x) \rightarrow x,y \vdash C(x,y)|C(y,x) \rightarrow \\
\rightarrow x,y,z \vdash C(x,y)|C(y,z)|C(z,x) \rightarrow \\
\rightarrow x,y,z,v \vdash C(x,y)|C(y,z)|C(z,v)|C(v,x)
\end{multline*}
In order to perform the reconfiguration into a star we need rules 
with nontrivial actions, like the second one. This allows to do:
\begin{multline*}
x,y,z,v \vdash C(x,y)|C(y,z)|C(z,v)|C(v,x) \xrightarrow[]{(x, r, \langle w \rangle),(y, r, \langle w \rangle),(z, r, \langle w \rangle),(v, r, \langle w \rangle)}\\
\rightarrow x,y,z,v,w \vdash S(x,w)|S(y,w)|S(z,w)|S(v,w)
\end{multline*}
Note that if an edge $C$ is rewritten into an edge $S$, then all the 
edges in the ring must use the same production, since they must 
synchronize via action $r$. They must agree also on $\n_\Lambda(x)$ 
for every $x$, thus all the newly created nodes are merged.
The whole transition is represented in Figure \ref{figure:reconf}.
\begin{figure}[!tb] \caption{Ring creation and reconfiguration to 
star} \label{figure:reconf}
\begin{center}
\resizebox{12,5cm}{!}{\includegraphics{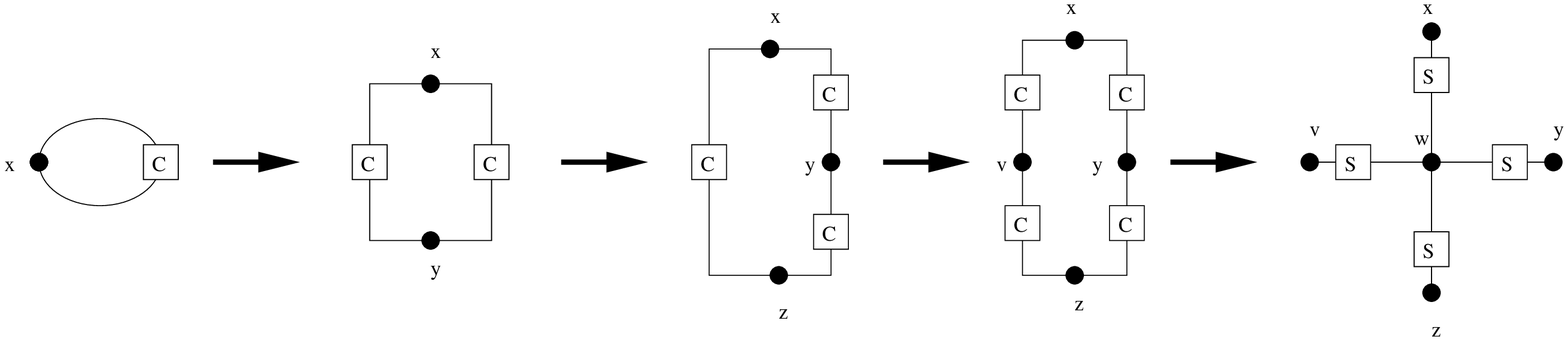}}
\end{center}
\end{figure}
\end{ex}

It is easy to show that if we can derive a transition $T$, then we can 
also derive every transition obtainable from $T$ by applying an injective 
renaming.

\begin{lemma}\label{lemma:injren}
Let $\mathcal{P}$ be a set of productions and $\sigma$ an injective 
substitution.\\
$\mathcal{P} \Vdash (\Gamma \vdash G \xrightarrow[]{\Lambda,\pi} \Phi 
\vdash G')$ iff:\\
$\mathcal{P} \Vdash (\Gamma\sigma \vdash G\sigma 
\xrightarrow[]{\Lambda',\pi'} \Phi\sigma \vdash G'\sigma)$\\
where $\Lambda'(x\sigma)=(\Lambda(x))\sigma$ and 
$x\sigma\pi'=x\pi\sigma$.
\end{lemma}
\begin{proof*}
By rule induction.
\end{proof*}

\subsection{Logic programming}\label{subsection:logicprog}
In this paper we are not interested in logic computations as refutations 
of goals for problem solving or artificial intelligence, but we consider 
logic programming \cite{lloydLP} as a \emph{goal rewriting mechanism}. We can consider 
logic subgoals as concurrent communicating processes that evolve according 
to the rules defined by the clauses and that use \emph{unification} as the 
fundamental interaction primitive. A presentation of this kind of use of 
logic programming can be found in \citeNS{bruniinteractive}.\\

In order to stress the similarities between logic programming and process 
calculi we present a semantics of logic programming based on a labelled 
transition system.

\begin{defin}\label{defin:plgrammar}
We have for clauses ($C$) and goals ($G$) the following grammar:

$$C \grammar A \leftarrow G$$ 
$$G \grammar G,G \  | \  A \ | \ \square$$ 

where $A$ is a logic atom, ``,'' is the AND conjunction and $\square$ is 
the empty goal.\\
We can assume ``,'' to be associative and commutative and with unit 
$\square$.
\end{defin}

The informal semantics of $A \leftarrow B_1,\dots,B_n$ is ``for every 
assignment of the variables, if $B_1,\dots,B_n$ are all true, then $A$ is 
true''.

A logic program is a set of clauses. Derivations in logic programming
are called SLD-derivations (from ``Linear resolution for Definite
clauses with Selection function''). We will also consider partial
SLD-derivations.

\begin{defin}[Partial SLD-derivation] \label{defin:SLD}
Let $P$ be a logic program.\\
We define a step of a SLD-resolution computation using the following 
rules:\\
$$\frac{H \leftarrow B_1,\dots,B_k \in P \qquad \theta=\mgu(\{A = H\rho\})}{P 
\Vdash A \xrightarrow[]{\theta} (B_1,\dots,B_k)\rho\theta} \qquad 
\textrm{atomic goal}$$
where $\rho$ is an injective renaming of variables such that all the 
variables in the clause variant $(H \leftarrow B_1,\dots,B_k)\rho$ are 
fresh.
$$\frac{P \Vdash G \xrightarrow[]{\theta} F}{P \Vdash G,G' 
\xrightarrow[]{\theta} F,G'\theta} \qquad \textrm{conjunctive goal}$$ 
We will omit $P \Vdash$ if $P$ is clear from the context.\\
A partial SLD-derivation of $P \cup \{G\}$ is a sequence (possibly 
empty) of steps of SLD-resolution allowed by program $P$ with initial goal 
$G$.
\end{defin}

\section{Mapping Fusion Calculus into Synchronized Hyperedge 
Replacement}\label{section:fusion2SHR} In this section we present a
mapping from Fusion Calculus to HSHR.\\ This mapping is quite complex
since there are many differences between the two formalisms. First of
all we need to bridge the gap between a process calculus and a graph
transformation formalism, and this is done by associating edges to
sequential processes and by connecting them according to the structure
of the system. Moreover we need to map Milner synchronization, which
is used in Fusion Calculus, into Hoare synchronization. In order to do
this we define some connection structures that we call amoeboids which
implement Milner synchronization using Hoare connectors. Since Hoare
synchronization involves all the edges attached to a node while Milner
one involves just pairs of connectors, we use amoeboids to force each
node to be shared by exactly two edges (one if the node is an
interface to the outside) since in that case the behaviour of Hoare
and Milner synchronization is similar. An amoeboid is essentially a
router (with no path-selection abilities) that connects an action
with the corresponding coaction. This is possible since in HSHR an
edge can do many synchronizations on different nodes at the same
time. Finally, some restrictions have to be imposed on HSHR in order
to have an interleaving behaviour as required by Fusion Calculus.\\

We define the translation on processes in the form $(\vec x)P$ where
$P$ is the parallel composition of sequential processes. Notice that
every process can be reduced to the above form by applying the
structural axioms: recursive definitions which are not inside a
sequential agent have to be unfolded once and scope operators which
are not inside a sequential agent must be taken to the outside.  We
define the translation also in the case $(\vec x)P$ is not closed \wrt
names (but it must be closed \wrt process variables) since this case
is needed for defining productions.\\

In the form $(\vec x)P$ we assume that the ordering of names in $(\vec x)$ 
is fixed, dictated by some structural condition on their occurrences in 
$P$.\\
For our purposes, it is also convenient to express process $P$ in $(\vec 
x)P$ as $P=P' \sigma$, where $P'$ is a linear agent, \ie every name in 
it appears once. We assume that the free names of $P'$ are fresh, namely 
$\fn(P') \cap \fn(P)=\emptyset$, and again structurally ordered. The 
corresponding vector is called $\fnarray(P')$.\\

The decomposition $P=P'\sigma$ highlights the role of amoeboids. In fact, 
in the translation, substitution $\sigma$ is made concrete by a graph 
consisting of amoeboids, which implement a router for every name in 
$\fn(P)$. More precisely, we assume the existence of edge labels 
$m_i$ and $n$ of ranks $i=2,3,\dots$ and $1$ respectively. Edges 
labelled by $m_i$ implement routers among $i$ nodes, while $n$ edges 
``close'' restricted names $\vec x$ in $(\vec x)P'\sigma$.\\

Finally, linear sequential processes $S$ in $P'$ must also be given a 
standard form. In fact, they will be modelled in the HSHR translation by 
edges labelled by $L_S$, namely by a label encapsulating $S$ itself. 
However in the derivatives of a recursive process the same sequential 
process can appear with different names an unbound number of times. To 
make the number of labels (and also of productions, as we will see in 
short) finite, for every given process, we choose standard names 
$x_1,\dots,x_n$ and order them structurally:\\
$S=\hat S(x_1,\dots,x_n) \rho_S$ with $S_1=S_2 \rho$ implying $\hat 
S_1=\hat S_2$ and $\rho_{S_1}=\rho\rho_{S_2}$.\\

We can now define the translation from Fusion Calculus to HSHR. The
translation is parametrized by the nodes in the vectors $\vec v$ and
$\vec w$ we choose to represent the names in $\vec x$ and
$\fnarray(P')$. We denote with $\bigpar_{x \in S} G_x$ the parallel
composition of graphs $G_x$ for each $x \in S$.
\begin{defin}[Translation from Fusion Calculus to 
HSHR]\label{defin:fc2hshr}
$\transl{(\vec x)P'\sigma}_{\vec v,\vec w}=\Gamma \vdash 
\transl{P'}\{\vec w/\fnarray(P')\}|\transl{\sigma}\{\vec v/\vec 
x\}\{\vec w/\fnarray(P')\}|\bigpar_{x \in \vec v} n(x)$\\
where: \\
$|\vec v|=|\vec x|$,\\
$|\vec w|=|\fnarray(P')|$,\\
$\vec v \cap \vec w=\emptyset$ \\
and with: \\
$\Gamma=\fn((\vec x)P'\sigma),\vec v,\vec w$.\\[2mm]
$\transl{0}=nil$\\ 
$\transl{S}=L_{\hat S}(x_1,\dots,x_n)\rho_S$ with $n=|\fn(\hat S)|$\\
$\transl{P_1|P_2}=\transl{P_1}|\transl{P_2}$\\[2mm]
$\transl{\sigma}=\bigpar_{x \in \im(\sigma)} m_{k+1}(x,\sigma^{-1}(x))$ where 
$k=|\sigma^{-1}(x)|$.
\end{defin}

In the above translation, graph $\transl{P'}$ consists of a set of 
disconnected edges, one for each sequential process of $(\vec 
x)P'\sigma$. The translation produces a graph with three kinds of 
nodes. The nodes of the first kind are those in $\vec w$. Each of them is 
adjacent to exactly two edges, one representing a sequential process 
of $P'$, and the other an amoeboid. Also the nodes in $\vec v$ are adjacent 
to two edges, an amoeboid and an $n$ edge. Finally the nodes in 
$\fn((\vec x)P'\sigma)$ are adjacent only to an amoeboid.\\

As mentioned above, translation $\transl{\sigma}$ builds an amoeboid for 
every free name $x$ of $P'\sigma$: it has $k+1$ tentacles, where $k$ are the 
occurrences of $x$ in $P'\sigma$, namely the free names of $P'$ mapped to 
it. Notice that the choice of the order within $\sigma^{-1}(x)$ is 
immaterial, since we will see that amoeboids are commutative \wrt their tentacles. However, to make the translation deterministic, 
$\sigma^{-1}(x)$ could be ordered according to some fixed precedence of 
the names.

\begin{ex}[Translation of a substitution]
Let $\sigma=\{y/x,y/z,r/w\}$. The translation of $\sigma$ is in Figure 
\ref{figure:amoeboids}.
\begin{figure}
\caption{Amoeboids for $\sigma$} \label{figure:amoeboids}
\resizebox{2cm}{!}{\includegraphics{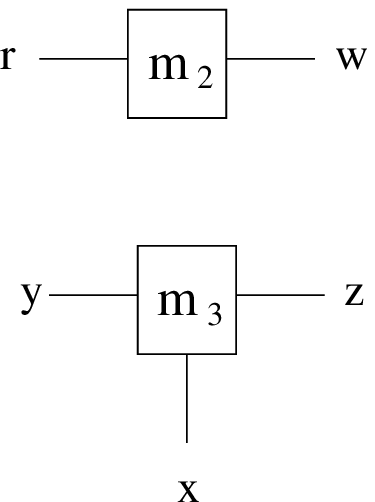}}
\end{figure}
\end{ex}

\begin{ex}[Translation of a process]\label{ex:processtr}
Let us consider the (closed) process $(uz)\overline{u}z.0|\rec 
X.(x)ux.(\overline{u}x.0|X)$. We can write it in the form $(\vec x) P$ 
as:\\
$(uzy) \overline{u}z.0|uy.(\overline{u}y.0|\rec 
X.(x)ux.(\overline{u}x.0|X))$\\
Furthermore we can decompose $P$ into $P'\sigma$ where:\\
$P'= \overline{u_1}z_1.0|u_2y_1.(\overline{u_3}y_2.0|\rec 
X.(x)u_4x.(\overline{u_5}x.0|X))$\\
$\sigma= \{ u/u_1,z/z_1,u/u_2,y/y_1,u/u_3,y/y_2,u/u_4,u/u_5 \}$.\\
We can now perform the translation.\\ 
We choose $\vec 
v=(u,z,y)$ and $\vec w=(u_1,z_1,u_2,y_1,u_3,y_2,u_4,u_5)$:
\begin{multline*}
\transl{(\vec x)P'\sigma}_{\vec v,\vec w}= 
u,z,y,u_1,z_1,u_2,y_1,u_3,y_2,u_4,u_5 \vdash \\ 
L_{\overline{x_1}x_2.0}(u_1,z_1)|L_{x_1x_2.(\overline{x_3}x_4.0|\rec 
X.(x)x_5x.(\overline{x_6}x.0|X))}(u_2,y_1,u_3,y_2,u_4,u_5)|\\m_6(u,u_1,u_2,u_3,u_4,u_5)|m_2(z,z_1)|m_3(y,y_1,y_2)|n(u)|n(z)|n(y)
\end{multline*}
\end{ex}

Now we define the productions used in the HSHR system.\\
We have two kinds of productions: auxiliary productions that are applied 
to amoeboid edges and process productions that are applied to process 
edges.\\
Before showing process productions we need to present the translation from 
Fusion Calculus prefixes into HSHR transition labels.

\begin{defin}\label{defin:acttransl}
The translation from Fusion Calculus prefixes into HSHR transition labels 
is the following:\\
$\transl{\alpha}=(\Lambda,\pi)$ where\\
if $\alpha=u\vec x$ then $\Lambda(u)=(in_n,\vec x)$, 
$\Lambda(x)=(\epsilon,\langle \rangle)$ if $x \neq u$ with $n=|\vec x|$, 
$\pi=id$\\
if $\alpha=\overline{u}\vec x$ then $\Lambda(u)=(out_n,\vec x)$, 
$\Lambda(x)=(\epsilon,\langle \rangle)$ if $x \neq u$ and $n=|\vec x|$, 
$\pi=id$\\
if $\alpha = \phi$ then $\Lambda=\Lambda_{\epsilon}$ and $\pi$ is any 
substitutive effect of $\phi$.\\
We will write $\transl{u\vec x}$ and $\transl{\overline{u}\vec x}$ as 
$(u,in_n,\vec x)$ and $(u,out_n,\vec 
x)$ respectively.
\end{defin}

\begin{defin}[Process productions]\label{defin:processprod}
We have a process production for each prefix at the top level of a linear 
standard sequential process (which has $\{x_1,\dots,x_n\}$ as free names).
Let $\sum_i \alpha_i.P_i$ be such a process. Its productions can be 
derived with the following 
inference rule:
$$\frac{\transl{P_j\xi}_{\vec v, \vec w}=\Gamma \vdash G \qquad
\transl{\alpha_j}=(\Lambda,\pi)}{x_1,\dots,x_n \vdash L_{\sum_i
\alpha_i.P_i}(x_1,\dots,x_n) \xrightarrow[]{\transl{\alpha_j}}
\Gamma,\Gamma' \vdash G|\transl{\xi}|\transl{\pi}|\bigpar_{x \in
\Gamma''} n(x)}$$ if $\vec v \cup \vec w$, $\{x_1,\dots,x_n\}$ and
$\fn(P_j)\xi$ are pairwise disjoint with $\xi$ injective renaming from
$\fn(P_j)$ to fresh names, $\Gamma'=x_1,\dots,x_n$ and
$\Gamma''=\Gamma' \setminus (\n(\Lambda) \cup \n(\pi) \cup \fn(P_j))$.
\end{defin}

We add some explanations on the derivable productions. Essentially, if
$\alpha_j.P_j$ is a possible choice, the edge labelled by the process
can have a transition labelled by $\transl{\alpha_j}$ to something
related to $\transl{P_j}_{\vec v,\vec w}$. We use $P_j\xi$ instead of
$P_j$ (and then we add the translation of $\xi$) to preserve the
parity of the number of amoeboid edges on each path (see Definition
\ref{defin:structured}).  The parameter $\vec v$ of the translation
contains fresh nodes for restricted names that are taken to the top
level during the normalization of $P_j$ while $\vec w$ contains the
free names in the normalization of $P_j$ (note that some of them may
be duplicated \wrt $P_j$, if this one contains recursion). If
$\alpha_j$ is a fusion $\phi$, according to the semantics of the
calculus, a substitutive effect $\pi$ of it should be applied to
$P_j$, and this is obtained by adding the amoeboids $\transl{\pi}$ in
parallel. Furthermore, $\Gamma \vdash G$ must be enriched in other two
ways: since nodes can never be erased, nodes which are present in the
sequential process, \ie the nodes in $\Gamma'$, must be added to
$\Gamma$. Also ``close'' $n$ edges must be associated to forgotten nodes
(to forbid further transitions on them and to have them connected to
exactly two edges in the result of the transition), provided they are
not exposed, \ie to nodes in $\Gamma ''$.\\

Note that when translating the RHS $(\vec x) P\sigma$ of productions
we may have names in $P\sigma$ which occur just once. Since they
are renamed by $\sigma$ and $\xi$, they will produce in the translation some
chains of $m_2$ connectors of even length, which, as we will see shortly, are behaviourally
equivalent to simple nodes. For simplicity, in the examples we will
use the equivalent productions where these connectors have been
removed and the nodes connected by them have been merged.

\begin{ex}[Translation of a production]\label{ex:productions}
Let us consider firstly the simple agent $\overline{x_1}x_2.0$.\\
The only production for this agent (where $\vec v=\vec w=\langle \rangle$) is:\\
$x_1,x_2 \vdash L_{\overline{x_1}x_2.0}(x_1,x_2) 
\xrightarrow[]{(x_1,out_1,\langle x_2 \rangle)} x_1,x_2 \vdash n(x_1)$\\
where we closed node $x_1$ but not node $x_2$ since the second one is 
exposed on $x_1$.\\
Let us consider a more complex example:\\
$x_1x_2.(\overline{x_3}x_4.0|\rec X. (x) x_5x.(\overline{x_6}x.0|X))$.\\
The process $\overline{x_3}x_4.0|\rec X. (x) x_5x.(\overline{x_6}x.0|X)\xi$ where $x_i\xi=x'_i$ 
can be transformed into:\\
$(y) \overline{y_1}y_2.0| y_3y_4.(\overline{y_5}y_6.0|\rec X. ((x) 
y_7x.(\overline{y_8}x.0|X)))\sigma$\\
where 
$\sigma=\{x'_3/y_1,x'_4/y_2,x'_5/y_3,y/y_4,x'_6/y_5,y/y_6,x'_5/y_7,x'_6/y_8\}$.\\
Its translation (with $\vec v=\langle y \rangle$ and $\vec w=\langle y_1,y_2,y_3,y_4,y_5,y_6,y_7,y_8 \rangle$) is:
\begin{multline*}
x'_3,x'_4,x'_5,x'_6,y_1,y_2,y_3,y_4,y_5,y_6,y_7,y_8,y 
\vdash\\L_{\overline{x_1}x_2.0}(y_1,y_2)| 
L_{x_1x_2.(\overline{x_3}x_4.0|\rec X. ((x) 
x_5x.(\overline{x_6}x.0|X)))}(y_3,y_4,y_5,y_6,y_7,y_8)|\\
m_2(x'_3,y_1)|m_2(x'_4,y_2)|m_3(x'_5,y_3,y_7)|m_3(x'_6,y_5,y_8)|m_3(y,y_4,y_6)|n(y)
\end{multline*}
Thus the production is:
\begin{multline*}
x_1,x_2,x_3,x_4,x_5,x_6 \vdash L_{x_1x_2.(\overline{x_3}x_4.0|\rec X. (x) 
x_5x.(\overline{x_6}x.0|X))}(x_1,x_2,x_3,x_4,x_5,x_6)\\ 
\xrightarrow[]{(x_1,in_1,\langle x_2 \rangle)}\\ 
x_1,x_2,x_3,x_4,x_5,x_6,y_3,y_4,y_5,y_6,y_7,y_8,y,x'_5,x'_6 \vdash\\ 
L_{\overline{x_1}x_2.0}(x_3,x_4)| L_{x_1x_2.(\overline{x_3}x_4.0|\rec X. 
((x) x_5x.(\overline{x_6}x.0|X)))}(y_3,y_4,y_5,y_6,y_7,y_8)|\\
m_3(x'_5,y_3,y_7)|m_3(x'_6,y_5,y_8)|m_3(y,y_4,y_6)|n(y)|m_2(x_5,x'_5)|m_2(x_6,x'_6)|n(x_1)
\end{multline*}
where for simplicity we collapsed $y_1$ with $x_3$ and $y_2$ with $x_4$.
\end{ex}

We will now show the productions for amoeboids. 
\begin{defin}[Auxiliary productions]
We have auxiliary productions of the form:
$$\Gamma \vdash m_k(\Gamma) \xrightarrow[]{(x_1,in_n,\vec
y_1),(x_2,out_n,\vec y_2)} \Gamma,\vec y_1,\vec y_2 \vdash m_k(\Gamma)
| \bigpar_{i=1 \dots |\vec y_1|} m_2(\vec y_1[i],\vec y_2[i])$$
We need such a production for each $k$ and $n$ and each pair of nodes
$x_1$ and $x_2$ in $\Gamma$ where $\Gamma$ is a chosen tuple of
distinct names with $k$ components and $\vec y_1$ and $\vec y_2$ are
two vectors of fresh names such that $|\vec y_1|=|\vec y_2|=n$.\\ 
Note
that we also have the analogous production where $x_1$ and $x_2$ are
swapped. In particular, the set of productions for a $m_k$ edge is
invariant \wrt permutations of the tentacles, modelling the
fact that its tentacles are essentially unordered.\\ 
We have no
productions for edges labelled with $n$, which thus forbid any
synchronization.
\end{defin}

The notion of amoeboid introduced previously is not sufficient for our
purposes. In fact, existing amoeboids can be connected using $m_2$
edges and nodes that are no more used can be closed using $n$
edges. Thus we present a more general definition of amoeboid for a set
of nodes and we show that, in the situations of interest, these
amoeboids behave exactly as the simpler $m_i$ edges.

\begin{defin}[Structured amoeboid]\label{defin:structured}
Given a vector of nodes $\vec s$, a structured amoeboid $M(\vec s)$
for the set of nodes $S$ containing all the nodes in $\vec s$ is any
connected graph composed by $m$ and $n$ edges that satisfies the
following properties:
\begin{itemize}
\item its set of nodes is of the form $S \cup I$, with $S \cap I=\emptyset$;
\item nodes in $S$ are connected to exactly one edge of the amoeboid;
\item nodes in $I$ are connected to exactly two edges of the amoeboid;
\item the number of edges composing each path connecting two distinct
nodes of $S$ is odd.
\end{itemize}
\end{defin}

Nodes in $S$ are called {\em external}, nodes in $I$ are called {\em
internal}. We consider {\em equivalent} all the amoeboids with the
same set $S$ of external nodes. The last condition is required since
each connector inverts the polarity of the synchronization, and we
want amoeboids to invert it.\\
Note that $m_{|S|}(\vec s)$ is an amoeboid for $S$.

\begin{lemma}\label{lemma:amoeboid}
If $M(\vec s)$ is a structured amoeboid for S, the
transitions for $M(\vec s)$ which are non idle and expose non
$\epsilon$ actions on at most two nodes $x_1,x_2 \in
S$ are of
the form:
$$S,I \vdash M(\vec s) \xrightarrow[]{\Lambda,id} S,I,I',\vec y_1,\vec
y_2 \vdash M(\vec s) | \bigpar_{i=1 \dots |\vec y_1|}
M(\vec y_1[i],\vec y_2[i]) | \bigpar \tilde M(\emptyset)$$ where
$\Lambda(x_1)=(in_n,\vec y_1)$ and $\Lambda(x_2)=(out_n,\vec y_2)$
(non trivial actions may be exposed also on some internal nodes) and
$\vec y_1$ and $\vec y_2$ are two vectors of fresh names such that
$|\vec y_1|=|\vec y_2|=n$.  Here $\tilde M(\emptyset)$ contains rings of
$m_2$ connectors connected only to fresh nodes which thus are
disconnected from the rest of the graph. We call them
\emph{pseudoamoeboids}.  Furthermore we have at least one transition
of this kind for each choice of $x_1$, $x_2$, $\vec y_1$ and $\vec
y_2$.
\end{lemma}
\begin{proof*}
See \ref{appendix:proofs}.
\end{proof*}

Thanks to the above result we will refer to structured amoeboids simply 
as amoeboids.

We can now present the results on the correctness and completeness of our 
translation.

\begin{theorem}[Correctness]\label{theorem:hoarecorrectness}
For each closed fusion process $P$ and each pair of vectors $\vec v$
and $\vec w$ satisfying the constraints of Definition
\ref{defin:fc2hshr}, if $P \rightarrow P'$ then there exist $\Lambda$,
$\Gamma$ and $G$ such that $\transl{P}_{\vec v, \vec w}
\xrightarrow[]{\Lambda,id} \Gamma \vdash G$. Furthermore $\Gamma
\vdash G$ is equal to $\transl{P'}_{\vec {v'}, \vec {w'}}$ (for some
$\vec {v'}$ and $\vec {w'}$) up to isolated nodes, up to injective
renamings, up to equivalence of amoeboids ($\Gamma \vdash G$ can have
a structured amoeboid where $\transl{P'}_{\vec {v'}, \vec {w'}}$ has a
simple one) and up to pseudoamoeboids.
\end{theorem}
\begin{proof*}
The proof is by rule induction on the reduction semantics.\\
See \ref{appendix:proofs}.
\end{proof*}

\begin{theorem}[Completeness]\label{theorem:hoarecompleteness}
For each closed fusion process $P$ and each pair of vectors $\vec v$
and $\vec w$ if $\transl{P}_{\vec v, \vec w}
\xrightarrow[]{\Lambda,\pi} \Gamma \vdash G$ with a HSHR transition
that uses exactly two productions for communication or one production
for a fusion action (plus any number of auxiliary productions) then $P
\rightarrow P'$ and $\Gamma \vdash G$ is equal to $\transl{P'}_{\vec
{v'}, \vec {w'}}$ (for some $\vec {v'}$ and $\vec {w'}$) up to
isolate nodes, up to injective renamings, up to equivalence of
amoeboids ($\Gamma \vdash G$ can have a structured amoeboid where
$\transl{P'}_{\vec {v'}, \vec {w'}}$ has a simple one) and up to
pseudoamoeboids.
\end{theorem}
\begin{proof*}
See \ref{appendix:proofs}.
\end{proof*}

These two theorems prove that the allowed transitions in the HSHR
setting correspond to reductions in the Fusion Calculus setting. Note
that in HSHR we must consider only transitions where we have either
two productions for communication or one production for a fusion
action. This is necessary to model the interleaving behaviour of
Fusion Calculus within the HSHR formalism, which is concurrent.  On
the contrary, one can consider the fusion equivalent of all the HSHR
transitions: these correspond to concurrent executions of many fusion
reductions. One can give a semantics for Fusion Calculus with that
behaviour. Anyway in that case the notion of equivalence of amoeboids
is no more valid, since different amoeboids allow different degrees of
concurrency. We thus need to constrain them. The simplest case is to
have only simple amoeboids, that is to have no concurrency inside a
single channel, but there is no way to force normalization of
amoeboids to happen before undesired transitions can occur. The
opposite case (all the processes can interact in pairs, also on the
same channel) can be realized, but it requires more complex auxiliary
productions.\\

Note that the differences between the final graph of a transition and
the translation of the final process of a Fusion Calculus reduction
are not important, since the two graphs have essentially the same
behaviours (see Lemma \ref{lemma:injren} for the effect of an
injective renaming and Lemma \ref{lemma:amoeboid} for the
characterization of the behaviour of a complex amoeboid; isolated
nodes and pseudoamoeboids are not relevant since different connected
components evolve independently). Thus the previous results can be
extended from transitions to whole computations.\\

Note that in the HSHR model the behavioural part of the system is 
represented by productions while the topological part is represented by 
graphs. Thus we have a convenient separation between the two 
different aspects.

\begin{ex}[Translation of a transition]\label{ex:fusion2hoare}
We will now show an example of the translation.
Let us consider the process:
$$(uxyzw)(Q(x,y,z)|\overline{u}xy.R(u,x)|uzw.S(z,w))$$
Note that it is already in the form $(\vec x)P$.
It can do the following transition:
\begin{multline*}
(uxyzw)(Q(x,y,z)|\overline{u}xy.R(u,x)|uzw.S(z,w)) \rightarrow\\
(uxy)(Q(x,y,z)|R(u,x)|S(z,w))\{x/z,y/w\}
\end{multline*}
We can write $P$ in the form:\\
$(Q(x_1,y_1,z_1)|\overline{u_1}x_2y_2.R(u_2,x_3)|u_3z_2w_1.S(z_3,w_2))\sigma$\\
where:\\ 
$\sigma=\{x/x_1,y/y_1,z/z_1,u/u_1,x/x_2,y/y_2,u/u_2,x/x_3,u/u_3,z/z_2,w/w_1,z/z_3,w/w_2\}$.\\

A translation of the starting process is:
\begin{multline*}
u,x,y,w,z,x_1,y_1,z_1,u_1,x_2,y_2,u_2,x_3,u_3,z_2,w_1,z_3,w_2 
\vdash\\
L_{Q(x_1,x_2,x_3)}(x_1,y_1,z_1)|L_{\overline{x_1}x_2x_3.R(x_4,x_5)}(u_1,x_2,y_2,u_2,x_3)|\\
L_{x_1x_2x_3.S(x_4,x_5)}(u_3,z_2,w_1,z_3,w_2)|m_4(u,u_1,u_2,u_3)|m_4(x,x_1,x_2,x_3)|\\
m_3(y,y_1,y_2)|m_4(z,z_1,z_2,z_3)|m_3(w,w_1,w_2)|n(u)|n(x)|n(y)|n(w)|n(z)
\end{multline*}
A graphical representation is in Figure \ref{figure:process1}.\\
\begin{figure}
\caption{$(uxyzw)(Q(x,y,z)|\overline{u}xy.R(u,x)|uzw.S(z,w))$} 
\label{figure:process1}
\resizebox{9cm}{!}{\includegraphics{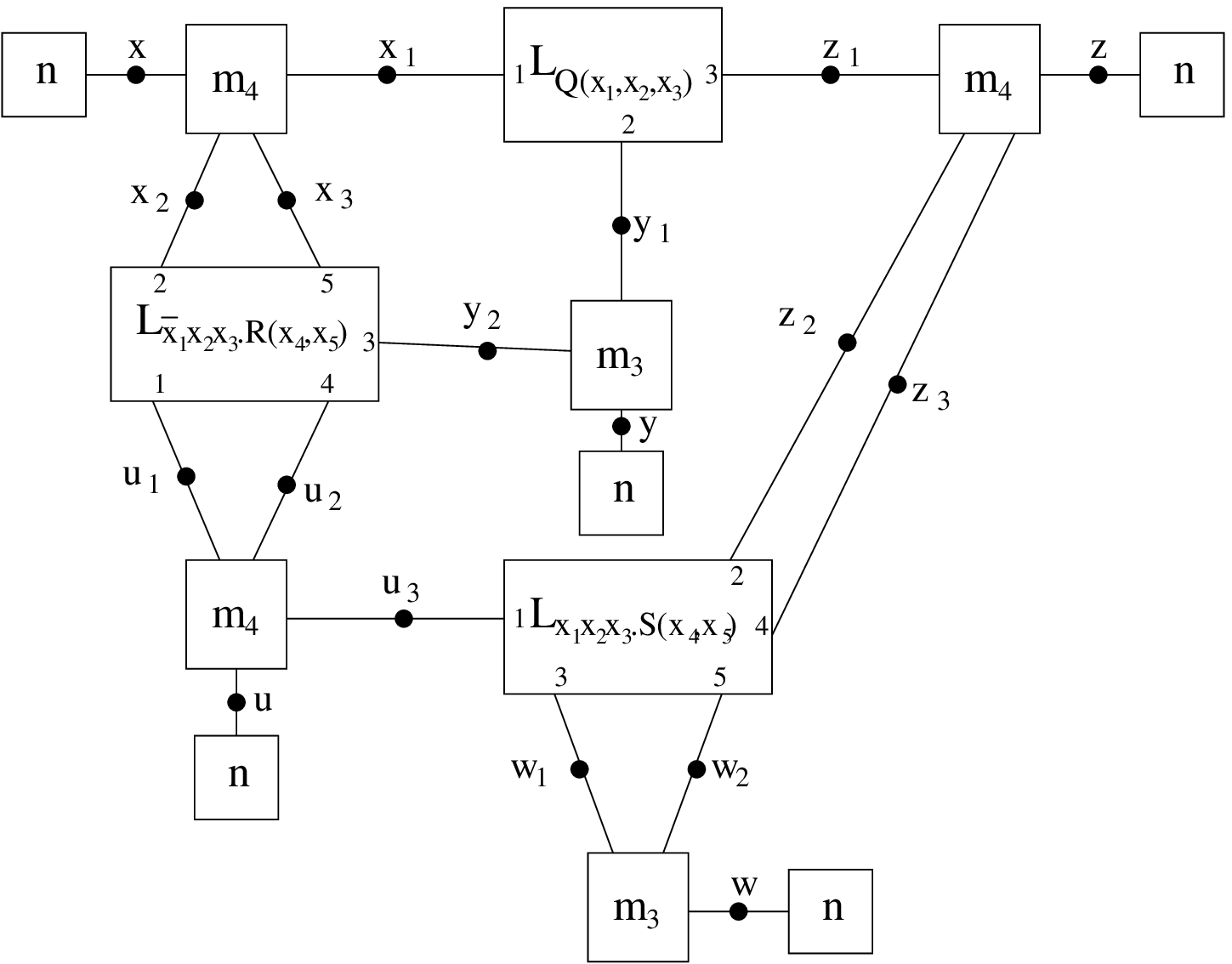}}
\end{figure}
We have the following process productions:
\begin{multline*}
y_1,y_2,y_3,y_4,y_5 \vdash 
L_{\overline{x_1}x_2x_3.R(x_4,x_5)}(y_1,y_2,y_3,y_4,y_5) 
\xrightarrow[]{(y_1,out_2,\langle y_2,y_3 \rangle)}\\ 
y_1,y_2,y_3,y_4,y_5 \vdash L_{R(x_1,x_2)}(y_4,y_5)|n(y_1)
\end{multline*}
\begin{multline*}
y_1,y_2,y_3,y_4,y_5 \vdash L_{x_1x_2x_3.S(x_4,x_5)}(y_1,y_2,y_3,y_4,y_5) 
\xrightarrow[]{(y_1,in_2,\langle y_2,y_3 \rangle)}\\ 
y_1,y_2,y_3,y_4,y_5 \vdash L_{S(x_1,x_2)}(y_4,y_5)|n(y_1)
\end{multline*}
In order to apply (suitable variants of) these two productions 
concurrently we have to synchronize their actions. This can be done since 
in the actual transition actions are exposed on nodes $u_1$ and $u_3$ 
respectively, which are connected to the same $m_4$ edge. Thus the 
synchronization can be performed (see Figure \ref{figure:process2}) and we 
obtain as final graph:
\begin{figure}
\caption{Graph with actions} \label{figure:process2}
\resizebox{9cm}{!}{\includegraphics{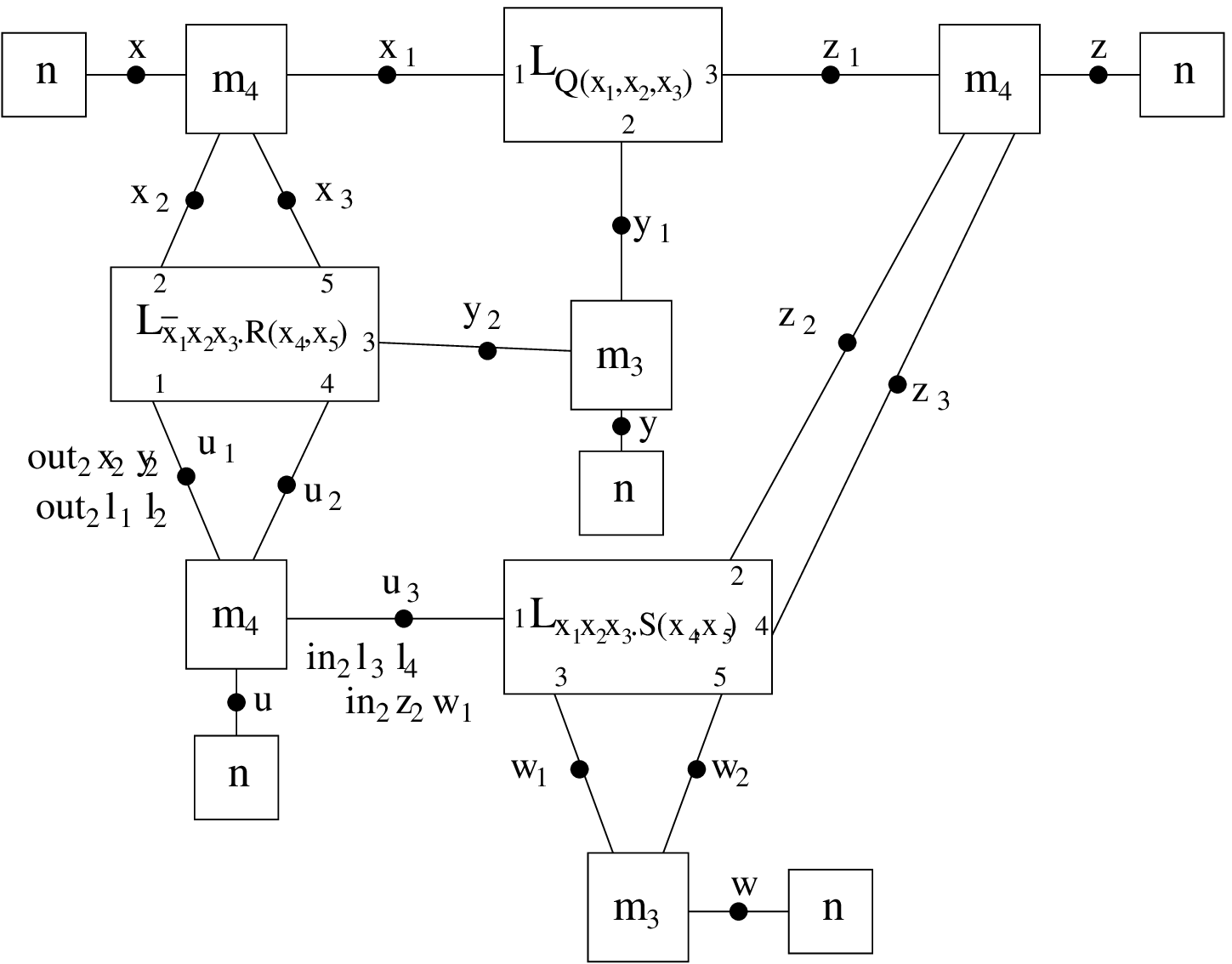}}
\end{figure}
\begin{multline*}
u,x,y,w,z,x_1,y_1,z_1,u_1,x_2,y_2,u_2,x_3,u_3,z_2,w_1,z_3,w_2 
\vdash\\
L_{Q(x_1,x_2,x_3)}(x_1,y_1,z_1)|L_{R(x_1,x_2)}(u_2,x_3)|n(u_1)|L_{S(x_1,x_2)}(z_3,w_2)|n(u_3)|\\
m_4(u,u_1,u_2,u_3)|m_4(x,x_1,x_2,x_3)|m_3(y,y_1,y_2)|m_4(z,z_1,z_2,z_3)|m_3(w,w_1,w_2)|\\
m_2(x_2,z_2)|m_2(y_2,w_1)|n(u)|n(x)|n(y)|n(w)|n(z)
\end{multline*}
which is represented in Figure \ref{figure:process3}.\\
\begin{figure}
\caption{Resulting graph} \label{figure:process3}
\resizebox{9cm}{!}{\includegraphics{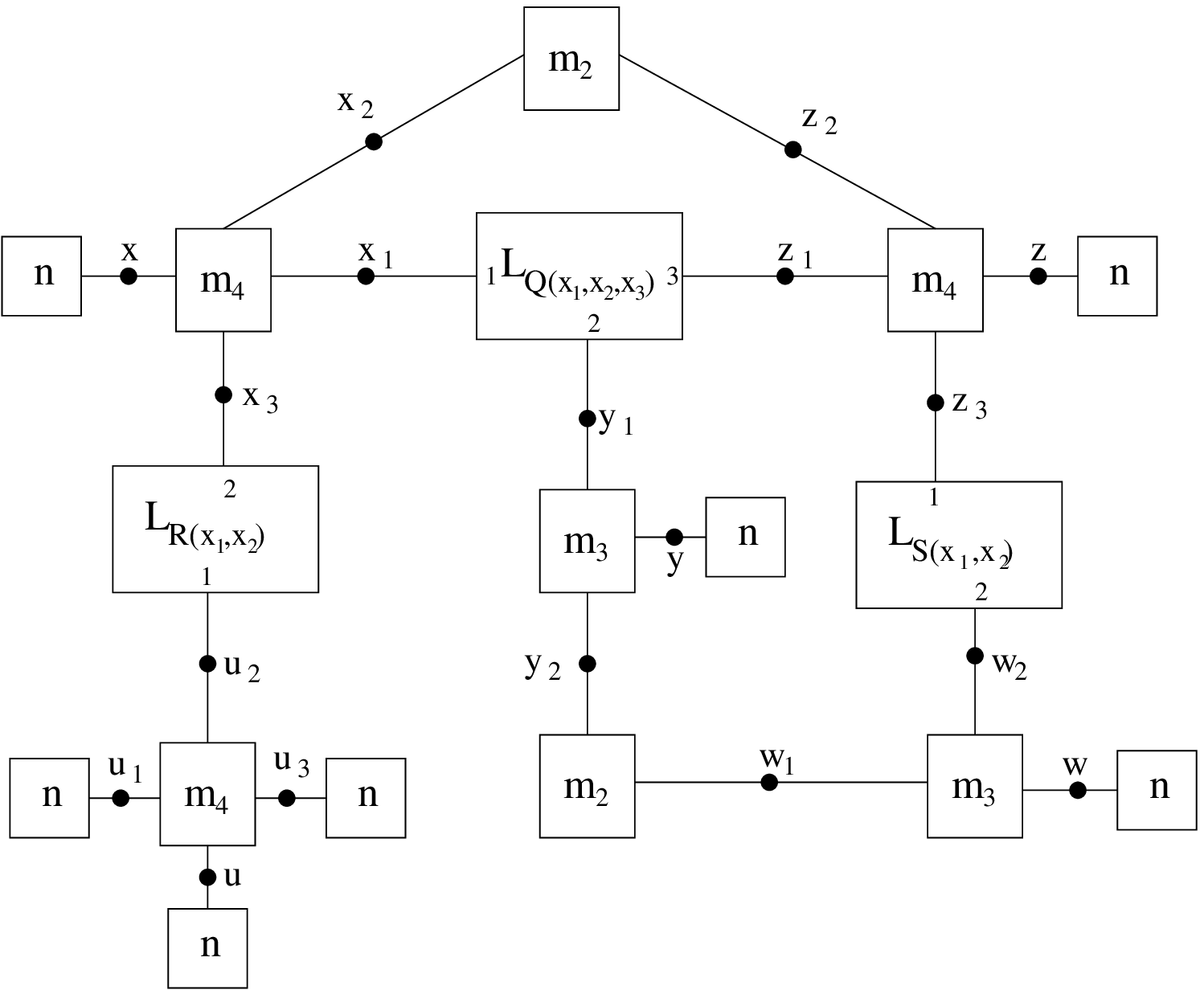}}
\end{figure}
The amoeboids connect the following tuples of nodes:\\ 
$(u,u_1,u_2,u_3)$, $(x,x_1,x_2,x_3,z_2,z,z_1,z_3)$, 
$(w,w_1,w_2,y_2,y,y_1)$. Thus, if we connect these sets of nodes with 
simple amoeboids instead of with complex ones, we have up to injective 
renamings a translation of $(uxy) Q(x,y,x)|R(u,x)|S(x,y)$ as required.
\end{ex}

\begin{ex}[Translation of a transition with recursion]
We will show here an example that uses recursion.
Let us consider the closed process $(uz) \overline{u}z|\rec X. (x) 
ux.(\overline{u}x.0|X)$. The translation of this process, as shown in 
Example \ref{ex:processtr} is:
\begin{multline*}
u,z,y,u_1,z_1,u_2,y_1,u_3,y_2,u_4,u_5 \vdash \\ 
L_{\overline{x_1}x_2.0}(u_1,z_1)|L_{x_1x_2.(\overline{x_3}x_4.0|\rec 
X.(x)x_5x.(\overline{x_6}x.0|X))}(u_2,y_1,u_3,y_2,u_4,u_5)|\\m_6(u,u_1,u_2,u_3,u_4,u_5)|m_2(z,z_1)|m_3(y,y_1,y_2)|n(u)|n(z)|n(y)
\end{multline*}

We need the productions for two sequential edges (for the first step): 
$\overline{x_1}x_2.0$ and $x_1x_2.(\overline{x_3}x_4.0|\rec X. (x) 
x_5x.(\overline{u}x_6.0|X))$.\\
The productions are the ones of Example \ref{ex:productions} (we write 
them here in a suitable $\alpha$-converted form):
\begin{multline*}
\hspace{2cm}u_1,z_1 \vdash L_{\overline{x_1}x_2.0}(u_1,z_1) 
\xrightarrow[]{(u_1,out_1,\langle z_1 \rangle)} u_1,z_1 \vdash n(u_1)\hspace{2cm}
\end{multline*}
\begin{multline*}
u_2,y_1,u_3,y_2,u_4,u_5 \vdash L_{x_1x_2.(\overline{x_3}x_4.0|\rec X. (x) 
x_5x.(\overline{x_6}x.0|X))}(u_2,y_1,u_3,y_2,u_4,u_5)\\ 
\xrightarrow[]{u_2,in_1,\langle y_1 \rangle}\\ 
u_2,y_1,u_3,y_2,u_4,u_5,w_1,w_2,w_3,w_4,w_5,w_6,y',u'_4,u'_5 \vdash\\ 
L_{\overline{x_1}x_2.0}(u_3,y_2)| L_{x_1x_2.(\overline{x_3}x_4.0|\rec X. 
((x) x_5x.(\overline{x_6}x.0|X)))}(w_1,w_2,w_3,w_4,w_5,w_6)|\\
m_3(u'_4,w_1,w_5)|m_3(u'_5,w_3,w_6)|m_3(y',w_2,w_4)|n(y')|m_2(u_4,u'_4)|m_2(u_5,u'_5)|n(u_2)
\end{multline*}
By using these two productions and a production for $m_6$ 
(the other edges stay idle) we have the following transition:
\begin{multline*}
u,z,y,u_1,z_1,u_2,y_1,u_3,y_2,u_4,u_5 \vdash \\ 
L_{\overline{x_1}x_2.0}(u_1,z_1)|L_{x_1x_2.(\overline{x_3}x_4.0|\rec 
X.(x)x_5x.(\overline{x_6}x.0|X))}(u_2,y_1,u_3,y_2,u_4,u_5)|\\m_6(u,u_1,u_2,u_3,u_4,u_5)|m_2(z,z_1)|m_3(y,y_1,y_2)|n(u)|n(z)|n(y)\\
\xrightarrow[]{(u_1,out_1,\langle z_1 \rangle)(u_2,in_1,\langle y_1 \rangle)}\\
u,y,z,x,u_1,z_1,u_2,y_1,u_3,y_2,u_4,u_5,w_1,w_2,w_3,w_4,w_5,w_6,y',u'_4,u'_5 
\vdash\\ n(u_1)|L_{\overline{x_1}x_2.0}(u_3,y_2)|
L_{x_1x_2.(\overline{x_3}x_4.0|\rec X. ((x) 
x_5x.(\overline{x_6}x.0|X)))}(w_1,w_2,w_3,w_4,w_5,w_6)|\\
m_3(u'_4,w_1,w_5)|m_3(u'_5,w_3,w_6)|m_3(y',w_2,w_4)|n(y')|m_2(u_4,u'_4)|m_2(u_5,u'_5)|n(u_2)|\\
m_6(u,u_1,u_2,u_3,u_4,u_5)|m_2(z_1,y_1)|m_2(z,z_1)|m_3(y,y_1,y_2)|n(u)|n(z)|n(y)
\end{multline*}
The resulting graph is, up to injective renaming and equivalence of amoeboids, a translation 
of:\\
$(uyy') (\overline{u}y.0| uy'.(\overline{u}y'.0|\rec X. (x) 
ux.(\overline{u}x.0|X)))$\\
as required.
\end{ex}


We end this section with a simple schema on the correspondence between the 
two models.\\[3mm]
\begin{oldtabular}{l|l|l|l}
\hline
Fusion & HSHR & Fusion & HSHR\\
\hline
Closed process & Graph & Reduction & Transition\\
Sequential process & Edge & Name & Amoeboid\\
Prefix execution & Production\hspace{5mm} & 0 & $Nil$\\
\hline
\end{oldtabular}

As shown in the table, we represent (closed) processes by graphs where
edges are sequential processes and amoeboids model names. The inactive
process $0$ is the empty graph $nil$. From a dynamic point of
view, Fusion reductions are modelled by HSHR transitions obtained
composing productions that represent prefix executions.

\section{Mapping Hoare SHR into logic programming}\label{section:HSHR2lp}
We will now present a mapping from HSHR into a subset of logic programming 
called Synchronized Logic Programming (SLP). The idea is to compose this 
mapping with the previous one obtaining a mapping from Fusion Calculus 
into logic programming.
\subsection{Synchronized Logic Programming}\label{subsection:SLP}
In this subsection we present Synchronized Logic Programming.\\
SLP has been introduced because logic programming allows for many
execution strategies and for complex interactions. Essentially SLP is
obtained from standard logic programming by adding a mechanism of
transactions. The approach is similar to the zero-safe nets approach
\cite{zero-safe} for Petri nets. In particular we consider that
function symbols are resources that can be used only inside a
transaction. A transaction can thus end only when the goal contains
just predicates and variables. During a transaction, which is called
big-step in this setting, each atom can be rewritten at most once. If
a transaction can not be terminated, then the computation is not
allowed. A computation is thus a sequence of big-steps.

This synchronized flavour of logic programming corresponds to HSHR since:
\begin{itemize}
\item used goals correspond to graphs (goal-graphs);
\item clauses in programs correspond to HSHR productions
(synchronized clauses);
\item resulting computations model HSHR computations (synchronized
computations).
\end{itemize}

\begin{defin}[Goal-graph] \label{defin:goalgraph}
We call goal-graph a goal which has no function symbols (constants are considered as functions of arity $0$). 
\end{defin}

\begin{defin}[Synchronized program] \label{defin:syncprograms}
A synchronized program is a finite set of synchronized rules, \ie 
definite program clauses such that:
\begin{itemize}
\item the body of each rule is a goal-graph;
\item the head of each rule is $A(t_1,\dots,t_n)$ where $t_i$ is
either a variable or a single function (of arity at least $1$) symbol
applied to variables.  If it is a variable then it also appears in the
body of the clause.
\end{itemize}
\end{defin}

\begin{ex} \label{ex:synchronized}\mbox{}\\
\begin{tabular}{cl}
$q(f(x),y) \leftarrow p(x,y)$ &
synchronized rule;\\
$q(f(x),y) \leftarrow p(x,f(y))$ &
not synchronized since $p(x,f(y))$ is not a goal-graph;\\
$q(g(f(x)),y) \leftarrow p(x,y)$ &
not synchronized since it contains nested functions;\\
$q(f(x),y,f(z)) \leftarrow p(x)$ &
\parbox[t]{8,3cm}{not synchronized since $y$ is an argument of the 
head predicate but it does not appear in the body;}\\
$q(f(x),f(z)) \leftarrow p(x)$ &
synchronized, even if $z$ does not appear in the body.
\end{tabular}
\end{ex}

In the mapping, the transaction mechanism is used to model the
synchronization of HSHR, where edges can be rewritten only if the
synchronization constraints are satisfied.  In particular, a clause
$A(t_1,\dots,t_n) \leftarrow B_1,\dots,B_n$ will represent a
production where the head predicate $A$ is the label of the edge in
the left hand side, and the body $B_1,\dots,B_n$ is the graph in the
right hand side. Term $t_i$ in the head represents the action
occurring in $x_i$, if $A(x_1,\dots,x_n)$ is the edge matched by the
production. Intuitively, the first condition of Definition
\ref{defin:syncprograms} says that the result of a local rewriting
must be a goal-graph. The second condition forbids synchronizations
with structured actions, which are not allowed in HSHR (this would
correspond to allow an action in a production to synchronize with a
sequence of actions from a computation of an adjacent
subgraph). Furthermore it imposes that we cannot disconnect from a
node without synchronizing on it \footnote{This condition has only the
technical meaning of making impossible some rewritings in which an
incorrect transition may not be forbidden because its only effect is
on the discarded variable.  Luckily, we can impose this condition
without altering the power of the formalism, because we can always
perform a special $foo$ action on the node we disconnect from and make
sure that all the other edges can freely do the same action.  For
example we can rewrite $q(f(x),y,f(z)) \leftarrow p(x)$ as
$q(f(x),foo(y),f(z)) \leftarrow p(x)$, which is an allowed
synchronized rule. An explicit translation of action $\epsilon$ can be
used too.}.

Now we will define the subset of computations we are interested in.

\begin{defin}[Synchronized Logic Programming]
Given a synchronized program $P$ we write:\\
$$G_1 \xRightarrow{\theta} G_2$$ 
iff $G_1 \xrightarrow[]{\theta'}\mstar G_2$ and all steps performed in the 
computation expand different atoms of $G_1$, $\theta'|_{\n(G_1)}=\theta$ 
and both $G_1$ and $G_2$ are 
goal-graphs.\\
We call $G_1 \xRightarrow{\theta} G_2$ a big-step and all the 
$\rightarrow$ 
steps in a big-step small-steps.\\
A SLP computation is:\\
$G_1 \Rightarrow^* G_2$
\ie a sequence of 0 or more big-steps.
\end{defin}

\subsection{The mapping}\label{subsection:HSHR2lp}

We want to use SLP to model HSHR systems. As a first step we need to
translate graphs, \ie syntactic judgements, to goals. In this
translation, edge labels are mapped into SLP predicates. Goals
corresponding to graphs will have no function symbols.  However
function symbols will be used to represent actions.  In the
translation we will lose the context $\Gamma$.

\begin{defin}[Translation for syntactic judgements]
We define the translation operator $\transl{-}$ as:

$$\transl{\Gamma \vdash L(x_1, \dots, x_n)} = L(x_1, \dots , x_n)$$
$$\transl{\Gamma \vdash G_1|G_2} = \transl{\Gamma \vdash 
G_1},\transl{\Gamma \vdash G_2}$$
$$\transl{\Gamma \vdash nil} = \square$$
\end{defin}

Sometimes we will omit the $\Gamma$ part of the syntactic
judgement. We can do this because it does not influence the
translation. For simplicity, we suppose that the set of nodes in the
SHR model coincides with the set of variables in SLP (otherwise we
need a bijective translation function). We do the same for edge labels
and names of predicates, and for actions and function symbols.

\begin{defin}
Let $\Gamma \vdash G$ and $\Gamma' \vdash G'$ be graphs.
We define the equivalence relation $\cong$ in the following way:
$\Gamma \vdash G \cong \Gamma' \vdash G'$ iff $G \equiv G'$.
\end{defin}

Observe that if two judgements are equivalent then they can be written 
as:\\
$\Gamma,\Gamma_{unused} \vdash G$\\
$\Gamma,\Gamma_{unused}' \vdash G$\\
where $\Gamma=\n(G)$.

\begin{theorem}[Correspondence of judgements and goal-graphs]
The operator $\transl{-}$ defines an isomorphism between judgements 
(defined up to $\cong$) and goal-graphs.
\end{theorem}
\begin{proof}
The proof is straightforward observing that the operator $\transl{-}$ 
defines a bijection between representatives of syntactic judgements and 
representatives of goal-graphs and the congruence on the two structures is 
essentially the same.
\end{proof}

We now define the translation from HSHR productions to definite clauses.

\begin{defin}[Translation from productions to clauses] 
\label{defin:tuostoruletranslation}
We define the translation operator $\transl{-}$ as:
$$\transl{L(x_1, \dots , x_n) \xrightarrow[]{\Lambda,\pi} G} = 
L(a_1(x_1\pi,\vec y_1), \dots , a_n(x_n\pi,\vec y_n)) \leftarrow \transl{G}$$
if $\Lambda(x_i) = (a_i,\vec y_i)$ for each $i \in \{1,\dots,n\}$ and if $a_i \neq 
\epsilon$. If $a_i = \epsilon$ we write simply $x_i\pi$ instead of 
$\epsilon(x_i\pi)$.
\end{defin}

The idea of the translation is that the condition given by an action 
$(x,a,\vec y)$ is represented by using the term $a(x\pi,\vec y)$ as 
argument in the position that corresponds to $x$. Notice that in this 
term $a$ is a function symbol and $\pi$ is a substitution.
During unification, 
$x$ will be bound to that term and, when other instances of $x$ are met, 
the corresponding term must contain the same function symbol (as required 
by Hoare synchronization) in order to be unifiable. Furthermore the 
corresponding tuples of transmitted nodes are unified. Since $x$ will 
disappear we need another variable to represent the node that corresponds to 
$x$. We use the first argument of $a$ to this purpose. If two nodes are 
merged by $\pi$ then their successors are the same as required.\\ 

Observe that we do not need to translate all the possible variants of the 
rules since variants with fresh variables are automatically built when the 
clauses are applied.
Notice also that the clauses we obtain are synchronized clauses.\\

The observable substitution contains information on $\Lambda$ and
$\pi$.  Thus given a transition we can associate to it a substitution
$\theta$. We have different choices for $\theta$ according to where we
map variables. In fact in HSHR nodes are mapped to their
representatives according to $\pi$, while, in SLP, $\theta$ cannot do
the same, since the variables of the clause variant must be all fresh.
The possible choices of fresh names for the variables change by an
injective renaming the result of the big-step.

\begin{defin}[Substitution associated to a transition] 
\label{defin:associated2}
Let $\Gamma \vdash G \xrightarrow[]{\Lambda,\pi} \Phi \vdash G'$ be a 
transition.
We say that the substitution $\theta_{\rho}$ associated to this transition 
is:\\
$\theta_{\rho}=\{(a(x\pi\rho,\vec y\rho)/x|\Lambda(x)=(a,\vec y),a \neq 
\epsilon\} \cup \{x\pi\rho/x\}|\Lambda(x)=(\epsilon,\langle \rangle)\}$\\
for some injective renaming $\rho$.
\end{defin}

We will now prove the correctness and the completeness of our translation.

\begin{theorem}[Correctness] \label{theorem:Tuostocorrectness}
Let $\mathcal{P}$ be a set of productions of a HSHR system as defined in 
definitions \ref{defin:production} and \ref{defin:hoarerules}. Let $P$ be 
the logic program obtained by translating the productions in $\mathcal{P}$ 
according to Definition \ref{defin:tuostoruletranslation}. If:\\
$\mathcal{P} \Vdash (\Gamma \vdash G \xrightarrow[]{\Lambda,\pi} \Phi 
\vdash G')$\\
then we can have in $P$ a big-step of Synchronized Logic Programming:\\
$\transl{\Gamma \vdash G} \xRightarrow{\theta_{\rho}}T$\\
for every $\rho$ such that $x\rho$ is a fresh variable unless possibly when $x \in 
\Gamma \land \Lambda(x)=(\epsilon,\langle \rangle)$. In that case we may have 
$x\rho=x$. Furthermore $\theta_\rho$ is associated to $\Gamma \vdash G 
\xrightarrow[]{\Lambda,\pi} \Phi \vdash G'$ and $T=\transl{\Phi 
\vdash G'}\rho$. Finally, 
used productions translate into the clauses used in the big-step and are applied to the 
edges that translate into the predicates rewritten by them.
\end{theorem}
\begin{proof*}
The proof is by rule induction.\\
See \ref{appendix:proofs}.
\end{proof*}

\begin{theorem}[Completeness] \label{theorem:Tuostocompleteness}
Let $\mathcal{P}$ be a set of productions of a HSHR system. Let $P$ be 
the logic program obtained by translating the productions in $\mathcal{P}$ 
according to Definition \ref{defin:tuostoruletranslation}. If we have in 
$P$ a big-step of logic programming:\\
$\transl{ \Gamma \vdash G } \xRightarrow{\theta}T$\\
then there exist $\rho$, $\theta'$, $\Lambda$, $\pi$, $\Phi$ and $G'$ such 
that $\theta=\theta'_{\rho}$ is associated to $\Gamma \vdash G 
\xrightarrow[]{\Lambda,\pi} \Phi \vdash G'$. Furthermore 
$T=\transl{ \Phi \vdash G' }\rho$ and $\mathcal{P} \Vdash (\Gamma 
\vdash G \xrightarrow[]{\Lambda,\pi} \Phi \vdash G')$. 
\end{theorem}
\begin{proof*}
See \ref{appendix:proofs}.
\end{proof*}

\begin{ex}\label{ex:fusion2lp}
We continue here Example \ref{ex:fusion2hoare} by showing how 
that fusion computation can be translated into a Synchronized Logic 
Programming computation.
\begin{multline*}
(uxyzw)(Q(x,y,z)|\overline{u}xy.R(u,x)|uzw.S(z,w)) \rightarrow\\
(uxy)(Q(x,y,z)|R(u,x)|S(z,w))\{x/z,y/w\}
\end{multline*}
Remember that a translation of the starting process is:
\begin{multline*}
u,x,y,w,z,x_1,y_1,z_1,u_1,x_2,y_2,u_2,x_3,u_3,z_2,w_1,z_3,w_2 
\vdash\\
L_{Q(x_1,x_2,x_3)}(x_1,y_1,z_1)|L_{\overline{x_1}x_2x_3.R(x_4,x_5)}(u_1,x_2,y_2,u_2,x_3)|\\
L_{x_1x_2x_3.S(x_4,x_5)}(u_3,z_2,w_1,z_3,w_2)|m_4(u,u_1,u_2,u_3)|m_4(x,x_1,x_2,x_3)|\\
m_3(y,y_1,y_2)|m_4(z,z_1,z_2,z_3)|m_3(w,w_1,w_2)|n(u)|n(x)|n(y)|n(w)|n(z)
\end{multline*}
We have the following productions:
\begin{multline*}
y_1,y_2,y_3,y_4,y_5 \vdash 
L_{\overline{x_1}x_2x_3.R(x_4,x_5)}(y_1,y_2,y_3,y_4,y_5) 
\xrightarrow[]{(y_1,out_2,\langle y_2,y_3 \rangle)}\\ 
y_1,y_2,y_3,y_4,y_5 \vdash L_{R(x_1,x_2)}(y_4,y_5)|n(y_1)
\end{multline*}
\begin{multline*}
y_1,y_2,y_3,y_4,y_5 \vdash L_{x_1x_2x_3.S(x_4,x_5)}(y_1,y_2,y_3,y_4,y_5) 
\xrightarrow[]{(y_1,in_2,\langle y_2,y_3 \rangle)}\\ 
y_1,y_2,y_3,y_4,y_5 \vdash L_{S(x_1,x_2)}(y_4,y_5)|n(y_1)
\end{multline*}
that corresponds to the clauses (we directly write suitably renamed 
variants):\\
$L_{\overline{x_1}x_2x_3.R(x_4,x_5)}(out_2(u'_1,x'_2,y'_2),x'_2,y'_2,u'_2,x'_3) 
\leftarrow L_{R(x_1,x_2)}(u'_2,x'_3)|n(u'_1)$\\
$L_{x_1x_2x_3.S(x_4,x_5)}(in_2(u'''_3,z'''_2,w'''_1),z'''_2,w'''_1,z'''_3,w'''_2) \leftarrow 
L_{S(x_1,x_2)}(z'''_3,w'''_2)|n(u'''_3)$\\
plus the clause obtained from the auxiliary production:
\begin{multline*}
m_4(u'',out_2(u''_1,x''_2,y''_2),u''_2,in_2(u''_3,z''_2,w''_1)) \leftarrow\\ 
m_4(u'',u''_1,u''_2,u''_3),m_2(x''_2,z''_2),m_2(y''_2,w''_1)
\end{multline*}
We obtain the big-step represented in Figure \ref{figure:bigstep}.
\begin{figure}
\caption{Big-step for a Fusion transition}
\label{figure:bigstep}
\begin{multline*}
L_{Q(x_1,x_2,x_3)}(x_1,y_1,z_1),L_{\overline{x_1}x_2x_3.R(x_4,x_5)}(u_1,x_2,y_2,u_2,x_3),\\
L_{x_1x_2x_3.S(x_4,x_5)}(u_3,z_2,w_1,z_3,w_2),\\
m_4(u,u_1,u_2,u_3),m_4(x,x_1,x_2,x_3),m_3(y,y_1,y_2),m_4(z,z_1,z_2,z_3),m_3(w,w_1,w_2),\\
n(u),n(x),n(y),n(w),n(z)\\
\xrightarrow[]{out_2(u'_1,x_2,y_2)/u_1,x_2/x'_2,y_2/y'_2,u_2/u'_2,x_3/x'_3}\\
L_{Q(x_1,x_2,x_3)}(x_1,y_1,z_1),L_{R(x_1,x_2)}(u_2,x_3),n(u'_1),
L_{x_1x_2x_3.S(x_4,x_5)}(u_3,z_2,w_1,z_3,w_2),\\
m_4(u,out_2(u'_1,x_2,y_2),u_2,u_3),m_4(x,x_1,x_2,x_3),m_3(y,y_1,y_2),\\
m_4(z,z_1,z_2,z_3),m_3(w,w_1,w_2),n(u),n(x),n(y),n(w),n(z)\\
\xrightarrow[]{u/u'',u'_1/u''_1,x_2/x''_2,y_2/y''_2,u_2/u''_2,in_2(u''_3,z''_2,w''_1)/u_3}\\
L_{Q(x_1,x_2,x_3)}(x_1,y_1,z_1),L_{R(x_1,x_2)}(u_2,x_3),n(u'_1),\\
L_{x_1x_2x_3.S(x_4,x_5)}(in_2(u''_3,z''_2,w''_1),z_2,w_1,z_3,w_2),\\
m_4(u,u'_1,u_2,u''_3),m_2(x_2,z''_2),m_2(y_2,w''_1),m_4(x,x_1,x_2,x_3),m_3(y,y_1,y_2),\\
m_4(z,z_1,z_2,z_3),m_3(w,w_1,w_2),n(u),n(x),n(y),n(w),n(z)\\
\xrightarrow[]{u''_3/u'''_3,z_2/z''_2,w_1/w''_1,z_2/z'''_2,w_1/w'''_1,z_3/z'''_3,w_2/w'''_2}\\
L_{Q(x_1,x_2,x_3)}(x_1,y_1,z_1),L_{R(x_1,x_2)}(u_2,x_3),n(u'_1),L_{S(x_1,x_2)}(z_3,w_2),n(u''_3),\\
m_4(u,u'_1,u_2,u''_3),m_2(x_2,z_2),m_2(y_2,w_1),m_4(x,x_1,x_2,x_3),m_3(y,y_1,y_2),\\
m_4(z,z_1,z_2,z_3),m_3(w,w_1,w_2),n(u),n(x),n(y),n(w),n(z)
\end{multline*}
\end{figure}
The observable substitution of the big-step is 
$\{out_2(u'_1,x_2,y_2)/u_1,in_2(u''_3,z_2,w_1)/u_3\}$. This is 
associated to the wanted HSHR transition with $\rho=\{u'_1/u_1,u''_3/u_3\}$ 
and by applying $\rho$ to the final graph of the HSHR transition we obtain:
\begin{multline*}
L_{Q(x_1,x_2,x_3)}(x_1,y_1,z_1)|L_{R(x_1,x_2)}(u_2,x_3)|n(u'_1)|L_{S(x_1,x_2)}(z_3,w_2)|n(u''_3)|\\
m_4(u,u'_1,u_2,u''_3)|m_4(x,x_1,x_2,x_3)|m_3(y,y_1,y_2)|m_4(z,z_1,z_2,z_3)|m_3(w,w_1,w_2)|\\
m_2(x_2,z_2)|m_2(y_2,w_1)|n(u)|n(x)|n(y)|n(w)|n(z)
\end{multline*}
that, translated, becomes the final goal of the big-step as required.
\end{ex}

We end this section with a simple schema on the correspondence between the 
two models.\\[3mm]
\begin{oldtabular}{l|l|l|l}
\hline
HSHR & SLP & HSHR & SLP\\
\hline
Graph & Goal & Transition & Big-step\\
Edge & Atomic goal & Node & Variable\\
Parallel comp. & And comp. & $Nil$ & $\square$\\
Production & Clause & Action & Function s.\\
\hline
\end{oldtabular}

Essentially the correspondence is given by the homomorphism between
graphs and goals, with edges mapped to atomic goals, nodes to
variables, parallel composition to And composition and $nil$ to
$\square$. Dynamically, HSHR transitions are modelled by big-steps,
that are transactional applications of clauses which model
productions. Finally, HSHR actions are modelled by function symbols.

\subsection{Using Prolog to implement Fusion 
Calculus}\label{subsection:meta}
The theorems seen in the previous sections can be used for implementation 
purposes. As far as Synchronized Logic Programming is concerned, in 
\citeNS{miatesi} a simple meta-interpreter is presented.

The idea is to use Prolog ability of dynamically changing the clause 
database to insert into it a set of clauses and a goal and to compute the 
possible synchronized computations of given length. This can be directly 
used to simulate HSHR transitions. In order to simulate Fusion Calculus 
processes we have to implement amoeboids using a bounded number of 
different connectors (note that $m_2$, $m_3$ and $n$ are enough) and to 
implement in the meta-interpreter the condition under which productions can 
be applied in a single big-step. This can be easily done. Furthermore this 
decreases the possible choices of applicable productions and thus improves the 
efficiency \wrt the general case.

\section{Conclusion}\label{section:conclusion}
In this paper we have analyzed the relationships between three different 
formalisms, namely Fusion Calculus, HSHR and logic programming.

The correspondence between HSHR and the chosen transactional version of logic 
programming, SLP, is complete and quite natural. Thus we can consider HSHR 
as a ``subcalculus'' of (synchronized) logic programming.

The mapping between Fusion Calculus and HSHR is instead more involved 
because it has to deal with many important differences:
\begin{itemize}
\item process calculi features vs graph transformation features;
\item interleaving models vs concurrent models;
\item Milner synchronization vs Hoare synchronization.
\end{itemize}
Hoare synchronization was necessary since our aim was to eventually
map Fusion Calculus to logic programming. If the aim is just to
compare Fusion Calculus and SHR it is possible to use SHR with Milner
synchronization, achieving a much simpler and complete mapping, which
considers the LTS of Fusion Calculus instead of reductions (see
\citeNP{ivancometa}).\\

We think that the present work can suggest several interesting lines of
development, dictated by the comparison of the three formalisms studied in
the paper. First, our implementation of routers in terms of
amoeboids is rather general and abstract, and shows that Fusion Calculus
names are a rather high level concept. They abstract out the behaviour of an
underlying network of connections which must be open and reconfigurable.
Had we chosen $\pi$-calculus instead (see a translation of $\pi$-calculus
to Milner SHR in \cite{hirschsynchronized}), we would have noticed
important differences. For instance, fusions are also considered in the
semantics of {\em open} $\pi$-calculus by Davide Sangiorgi \cite{openpi}, but in that
work not all the names can be fused: newly extruded names cannot be merged
with previously generated names. This is essential for specifying nonces
and session keys for secure protocols. Instead, Fusion Calculus does not
provide equivalent constructs. Looking at our translation, we can conclude
that logic programming does not offer this feature, either. Thus logic
programming is a suitable counterpart of Fusion Calculus, but it should be
properly extended for matching open $\pi$-calculus and security
applications.

In a similar line of thought, we observe that we have a scope
restriction operator in the Fusion Calculus, but no restriction is
found in our version of HSHR. We think this omission simplifies our
development, since no restriction exists in ordinary logic
programming, either. However versions of SHR with restriction have
been considered \cite{hirschsynchronized,tuostoambients,miatesi}. Also
(synchronized) logic programming can be smoothly extended with a
restriction operator \cite{miatesi}. More importantly, Fusion Calculus
is equipped with an observational abstract semantics based on (hyper)
bisimulation. We did not consider a similar concept for SHR or logic
programming, since we considered it outside the scope of the
paper. Furthermore our operational correspondence between HSHR and SLP
is very strong and it should respect any reasonable abstract
semantics. The mapping from Fusion Calculus into HSHR deals only with
closed terms, thus no observations can be considered. However a
bisimulation semantics of SHR has been considered in
\cite{konigobservational}, and an observational semantics of logic
programming is discussed in \cite{bruniinteractive}.

Another comment concerns concurrency. To prove the equivalence of Fusion
Calculus and of its translation into HSHR we had to restrict the possible
computations of the latter. On the contrary, if all computations
were allowed, the same translation would yield a concurrent semantics of
Fusion Calculus, that we think is worth studying. For instance in the presence
of concurrent computations not all equivalent amoeboids would have the
same behaviour, since some of them would allow for more parallelism than
others.

Finally we would like to emphasize some practical implication of our work.
In fact, logic programming is not only a model of computation, but also a
well developed programming paradigm. Following the lines of our
translation, implementations of languages based on Fusion Calculus and HSHR
could be designed, allowing to exploit existing ideas, algorithms and tools
developed for logic programming. 

\bibliographystyle{acmtrans}

\appendix

\section{Proofs}\label{appendix:proofs}
We have here the proofs that are missing in the main part and some lemmas 
used in these proofs. Lemmas are just before the proofs that use them.

\begin{proof}[Proof of Lemma \ref{lemma:amoeboid}]
Notice that all the auxiliary productions perform two non trivial
actions, and because of Hoare synchronization and because each node is
shared by at most two edges, synchronizing edges form chains. There
are two alternatives: each chain either starts and begins on an
external node, or it is a cycle that contains only internal nodes.
Exactly one chain must be of the first type. In fact if we have no
chain of that kind we have only trivial actions on nodes in
$S$. Also, if we have more than one, we have more than two non trivial
actions on nodes in $S$. Thanks to the last condition of Definition
\ref{defin:structured} this chain contains an odd number of
connectors, which must be $m$ connectors. One can easily check by
induction on the (odd) length of the chain that the transition creates
an amoeboid $M(\vec y_1[i],\vec y_2[i])$ for each component of vectors $\vec
y_1$ and $\vec y_2$ of fresh names and that these vectors are exposed
on the external nodes together with an $in_n$ and an $out_n$ action,
where $n=|\vec y_1|=|\vec y_2|$.

Let us now consider the other kind of chains: these chains produce
rings of $m_2$ edges connected only to fresh nodes, which thus
correspond to isolate subgraphs. Furthermore, they affect only the
labels of internal nodes as required.
\end{proof}

\begin{lemma}\label{lemma:amoeboidconn}
Given a set of amoeboids for $\sigma$ and a substitutive effect
$\theta$ of a fusion $\phi=\{x_i=y_i|i=1 \dots n\}$ then
$\transl{\sigma}|\bigpar_{i=1 \dots n} M(x_i,y_i)$ is a set of
amoeboids for $\sigma\theta$.
\end{lemma}
\begin{proof}
Since we are working up to injective renamings we only have to prove
that two names are connected by $\transl{\sigma}|\bigpar_{i=1 \dots n}
M(x_i,y_i)$ iff they are merged by $\sigma\theta$ and that all the
paths connecting external nodes in the final graph have odd length. By
definition two names are connected by $\transl{\sigma}$ iff they are
merged by $\sigma$. Assume that two names $x$ and $y$ are merged by
$\sigma\theta$. Then their images along $\sigma$ are merged by
$\theta$.  This means that we have in $\phi$ a chain of equalities
from $x\sigma$ to $y\sigma$. Thus we have amoeboids connecting the
amoeboids of $x\sigma$ and of $y\sigma$, thus $x$ and $y$ are
connected. Assume now that $x$ and $y$ are connected. Then there
exists an amoeboid connecting the amoeboids for $x\sigma$ and
for $y\sigma$. Thus $\theta$ merges $x\sigma$ and $y\sigma$ as
required. Finally note that all the paths between external nodes are
created by connecting existing paths via new amoeboids. In particular,
each new path is composed by $n$ old paths and $n-1$ new
amoeboids. Thus its length is the sum of $2n-1$ odd lengths and thus
it is odd.
\end{proof}

\begin{proof}[Proof of Theorem \ref{theorem:hoarecorrectness}]
The proof is by rule induction on the reduction semantics.\\
Let us consider the reduction rule:
$$(\vec z) (R|(\dots + u \vec x .P)|(\overline{u} \vec y.Q+\dots)) 
\rightarrow (\vec z)(R|P|Q)\sigma$$
where $|\vec x|=|\vec y|$ and $\sigma$ is a substitutive effect of $\{ 
\vec x = \vec y \}$ such that $\dom(\sigma) \subseteq \vec z$.

In order for that process to be in the standard form we just need to
make $R$ sequential by unfolding recursion and taking bound names to
the outside if required. For simplicity, we show just the case where
$R$ is already sequential, the other being essentially equal.

Let $P_1$ be $(\dots + u \vec x .P)$ and $Q_1$ be $(\overline{u} \vec y.Q+\dots)$. 

The translation of the LHS has the form:
$$\Gamma \vdash L_{\hat R'}(\vec r)|L_{\hat P'_1}(\vec p_1)|L_{\hat Q'_1}(\vec q_1)|\transl{\theta}|\bigpar_{x \in \vec v} n(x)$$
where $R'=\hat R'\rho_{R'}$, $P'_1=\hat P'_1\rho_{P'_1}$, $Q'_1=\hat
Q'_1\rho_{Q'_1}$ and $R'\theta=R$, $P'_1\theta=P_1$, $Q'_1\theta=Q_1$.
We also use $(u'_1 \vec x'.P')\theta=u\vec x.P$ and $(u'_2 \vec y'.Q')\theta=u\vec y.Q$.

We have for $L_{\hat P'_1}$ and $L_{\hat Q'_1}$ the two following productions:

$$\vec p_1 \vdash L_{\hat P'_1}(\vec p_1) \xrightarrow[]{(u'_1,in_{|\vec x'|},\vec x')} \vec p_1,\vec x',\vec v_P \vdash \transl{P}_{\vec v_p,\vec w_P}|\bigpar_{x \in \Gamma''_P} n(x)$$
$$\vec q_1 \vdash L_{\hat Q'_1}(\vec q_1) \xrightarrow[]{(u'_2,out_{|\vec y'|},\vec y')} \vec q_1,\vec y',\vec v_Q \vdash \transl{Q}_{\vec v_Q,\vec w_Q}|\bigpar_{x \in \Gamma''_Q} n(x)$$

The choice of the parameters $\vec v_P$ and $\vec v_Q$ is not
important since they are fresh names and we work up to injective
renamings. The choice of $\vec w_P$ and $\vec w_Q$ is important
instead, since they have to correspond to the nodes in the LHS that
are still used by the process. The choice performed in Definition
\ref{defin:processprod} ensures that.

Note that $\transl{\theta}$ contains an amoeboid for a set $U$ with
$u'_1,u'_2 \in U$. Thus from Lemma \ref{lemma:amoeboid} the two above
productions can synchronize via the amoeboid. Using rule (idle) for
the other edges, we obtain as a result:

\begin{multline*}
\Gamma, \Gamma_{Int} \vdash L_{\hat R'}(\vec r)|\transl{P}_{\vec
v_P,\vec w_P}|\bigpar_{x \in \Gamma''_P} n(x)|\transl{Q}_{\vec
v_Q,\vec w_Q}|\bigpar_{x \in \Gamma''_Q}
n(x)|\transl{\theta}|\bigpar_{x \in \vec v} n(x)| \\ \bigpar_{i=1 \dots
|\vec x'|} M(\vec x'[i],\vec y'[i]) | \bigpar \tilde M(\emptyset)
\end{multline*}

Note that the fusion $\pi$ of the transition is the identity since
each synchronization involves two edges, and at least one of them is
an amoeboid which exposes fresh names. Furthermore amoeboids expose
each fresh name twice. Thus no names in $\Gamma$ are merged.

Also, in the final graph each node is shared by two edges. In fact,
the only connections whose cardinality is not preserved by productions are the
ones between nodes whose references are exposed, namely $\vec x'$ and
$\vec y'$, and the corresponding process edges, but these nodes are
connected to the new $M$ amoeboids created by the auxiliary
productions.

In particular, $\bigpar_{x \in \Gamma''_P} n(x)|\bigpar_{x \in
\Gamma''_Q} n(x)|\transl{\theta}$ is an amoeboid for
$\theta|_{\fn(R'|P'|Q')}$. Furthermore thanks to Lemma
\ref{lemma:amoeboidconn} by adding in parallel $\bigpar_{i=1 \dots
|\vec x'|} M(\vec x'[i],\vec y'[i])$ we obtain an amoeboid for
$\theta|_{\fn(R'|P'|Q')}\sigma$.

Thus we can rewrite the final graph up to pseudoamoeboids as:

$$\Gamma, \Gamma_{Int} \vdash L_{\hat R'}(\vec r)|\transl{P}_{\vec
v_P,\vec w_P}|\transl{Q}_{\vec v_Q,\vec
w_Q}|\transl{\theta|_{\fn(R'|P'|Q')}\sigma}|\bigpar_{x \in \vec v}
n(x)$$

The RHS of the fusion rule is $(\vec z)(R|P|Q)\sigma$, which has to be
normalized into $(\vec z \vec v_P \vec v_Q)(R|\tilde P|\tilde
Q)\theta_1\sigma$ with $(\vec v_P)\tilde P \equiv P$ and $(\vec
v_Q)\tilde Q \equiv Q$. Furthermore we can choose the names (since we
are reasoning up to injective renamings) in such a way that
$\theta_1=\theta|_{\fn(R'|P'|Q')}$ (plus an injective renaming on
names in $\vec v_p$ and $\vec v_Q$ which corresponds to an equivalence
on the resulting amoeboids).

Thus the translation of the RHS is equivalent to:

$$\Gamma, \Gamma_{Int'} \vdash L_{\hat R'}(\vec r)|\transl{P}_{\vec v_P,\vec
w_P}|\transl{Q}_{\vec v_Q,\vec w_Q}|\transl{\theta_1\sigma}|\bigpar_{x
\in \vec v} n(x)$$

The correctness of the rule follows.

Let us consider now rule:
$$(\vec z) (R|(\dots + \phi .P)) \rightarrow (\vec z)(R|P)\sigma$$
where $\sigma$ is a substitutive effect of $\phi$ such that $\dom(\sigma) \subseteq \vec z$.

We use essentially the same technique as before. Let us suppose $R$
already sequential. Let $P_1$ be $\dots+\phi.P$.  The translation of
the LHS has the form:

$$\Gamma \vdash L_{\hat R'}(\vec r)|L_{\hat P'_1}(\vec p_1)|\transl{\theta}|\bigpar_{x \in \vec v} n(x)$$
where $R'=\hat R'\rho_{R'}$, $P'_1=\hat P'_1\rho_{P'_1}$ and $R'\theta=R$, $P'_1\theta=P_1$.
We also use $(\phi'.P')\theta=\phi.P$.

We have for $L_{\hat P'_1}$ the following production:

$$\vec p_1
\vdash L_{\hat P'_1}(\vec p_1) 
\xrightarrow[]{\Lambda_{\epsilon}} \vec p_1,\vec v_P \vdash 
\transl{P}_{\vec v_P,\vec w_P}|\transl{\sigma}|\bigpar_{x \in \Gamma''} n(x)$$

For $\vec v_P$ and $\vec w_P$ the considerations for the preceding rule are still valid.

Using rule (idle) for the other edges, we obtain as a result:

$$\Gamma, \Gamma_{Int} \vdash L_{\hat R'}(\vec r)|\transl{P}_{\vec
v_P,\vec w_P}|\transl{\sigma}|\bigpar_{x \in \Gamma''_P} n(x)|\transl{\theta}|\bigpar_{x \in \vec v} n(x)$$

Note that the fusion part of the transition label is an identity since
we have only trivial synchronizations. 

Also, in the final graph each node is shared by two edges since the
production preserves the cardinality of connected edges for each node.

In particular, $\bigpar_{x \in \Gamma''_P} n(x)|\transl{\theta}$ is an amoeboid for
$\theta|_{\fn(R'|P')}$. Furthermore, note that $\transl{\sigma}$ has the form $\bigpar_{i=1 \dots
n} m_2(x_{i},y_{i})$ and thus thanks to Lemma
\ref{lemma:amoeboidconn} by adding it in parallel we obtain an amoeboid for
$\theta|_{\fn(R'|P')}\sigma$.

Thus we can rewrite the final graph as:

$$\Gamma, \Gamma_{Int} \vdash L_{\hat R'}(\vec r)|\transl{P}_{\vec
v_P,\vec w_P}|\transl{\theta|_{\fn(R'|P')}\sigma}|\bigpar_{x \in \vec v}
n(x)$$

The RHS of the fusion rule is $(\vec z)(R|P)\sigma$, which has to be
normalized into $(\vec z \vec v_P)(R|\tilde P)\theta_1\sigma$ with $(\vec v_P)\tilde P \equiv P$. Furthermore we can choose the names (since we
are reasoning up to injective renamings) in such a way that
$\theta_1=\theta|_{\fn(R'|P')}$ (plus an injective renaming on
names in $\vec v_P$ which corresponds to an equivalence
on the resulting amoeboids).

Thus the translation of the RHS is equivalent to:

$$\Gamma, \Gamma_{Int'} \vdash L_{\hat R'}(\vec r)|\transl{P}_{\vec v_P,\vec
w_P}|\transl{\theta_1\sigma}|\bigpar_{x
\in \vec v} n(x)$$

The correctness of the rule follows.

Consider now the rule:

$$\frac{P \equiv P', P' \rightarrow Q' , Q' \equiv Q}{P \rightarrow Q}$$
Equivalent agents are converted into the same representative (up to 
$\alpha$-conversion) before being translated, thus the translation of $P$ 
and $P'$ are equal up to injective renamings. Similarly 
for the translation of $Q$ 
and $Q'$, thus the thesis 
follows.
\end{proof}

\begin{lemma}\label{lemma:mgusubst}
Let $\theta_1$ and $\theta_2$ be idempotent substitutions. Let
$\eqn(\theta_1)=\{x=y|x/y \in \theta_1\}$.  Then $\mgu(\eqn(\theta_1)
\cup \eqn(\theta_2))=\theta_1\mgu(\eqn(\theta_2)\theta_1)$.
\end{lemma}
\begin{proof}
See \citeN{palamidessialg}.
\end{proof}

\begin{lemma}\label{lemma:det}
Given a graph $\Gamma \vdash G$ and one or zero productions for each edge 
in $G$ let:\\ 
$\Gamma \vdash G \xrightarrow[]{\Lambda_1,\pi_1} \Phi_1 \vdash G'_1$\\
$\Gamma \vdash G \xrightarrow[]{\Lambda_2,\pi_2} \Phi_2 \vdash G'_2$\\
be two transitions obtained by applying the chosen production for each 
edge (and using the (idle) rule if no production is chosen).
Then there exists an injective renaming $\sigma$ such that:
\begin{itemize}
\item $\Lambda_1(x)=\sigma(\Lambda_2(x))$ if $x$ is not an isolated node;
\item $\pi_1=\sigma\pi_2$;
\item $\Phi_1=\sigma(\Phi_2)$;
\item $G'_1=\sigma(G'_2)$.
\end{itemize}
\end{lemma}

\begin{proof}
The proof is a simple rule induction if one proves that derivations
can be done in a standard way, namely by applying to the axioms first
rules (par), then rule (merge) and finally rules (new). This can be
proved by showing that one can exchange the order of rules and that
one can substitute two applications of (merge) with substitutions
$\sigma$ and $\sigma'$ with just one application with substitution
$\sigma'\sigma$.  We have many cases to consider, but they are not so
interesting. As examples we will show the detailed proof for
commutation of rule (merge) and (par) and for composition of two
different rules (merge).

Let us consider the first case. Suppose we have a part of a derivation
of the form:

\begin{equation}\label{eqn:merge-par}
\infer{\Gamma\sigma, \Gamma' \vdash G_1\sigma|G'_1 \xrightarrow[]{\Lambda_1 \cup \Lambda', \pi_1 \cup \pi'} \Phi_1, \Phi' \vdash G_2\sigma\rho|G'_2}
   {\infer
       {\Gamma\sigma \vdash G_1\sigma \xrightarrow[]{\Lambda_1, \pi_1} \Phi_1 \vdash G_2\sigma\rho} 
       {\Gamma \vdash G_1 \xrightarrow[]{\Lambda, \pi} \Phi \vdash G_2}
   & \Gamma' \vdash G'_1 \xrightarrow[]{\Lambda', \pi'} \Phi' \vdash G'_2
   }
\end{equation}

where for readability we have not written explicitly the side
conditions (see Definition \ref{defin:hoarerules}).  Then we must also
have a derivation for the same transition obtained applying rule (par)
first and then rule (merge):

\begin{equation}\label{eqn:par-merge}
\infer{(\Gamma, \Gamma')\sigma \vdash (G_1|G'_1)\sigma \xrightarrow[]{\Lambda'', \pi''} \Phi'' \vdash (G_2|G'_2)\sigma\rho'}
   {\infer{\Gamma, \Gamma' \vdash G_1|G'_1 \xrightarrow[]{\Lambda \cup \Lambda', \pi \cup \pi'} \Phi, \Phi' \vdash G_2|G'_2}
        {\Gamma \vdash G_1 \xrightarrow[]{\Lambda, \pi} \Phi \vdash G_2 & \Gamma' \vdash G'_1 \xrightarrow[]{\Lambda', \pi'} \Phi' \vdash G'_2}
   }
\end{equation}

We have to prove that this derivation is allowed and that the
resulting transition is the one derived also by derivation
\ref{eqn:merge-par}. The first step is allowed iff $(\Gamma \cup \Phi)
\cap (\Gamma' \cup \Phi')=\emptyset$. Since the first derivation is
allowed by hypothesis, then $(\Gamma\sigma \cup \Phi_1) \cap (\Gamma'
\cup \Phi')=\emptyset$. Thus the only problem is when a name which is
after renamed (by $\sigma$ or $\rho$) creates a conflict. In that case
thanks to Lemma \ref{lemma:injren} we can suppose to start with a
different name. The final result of the derivation is not changed by
that since the name disappears. Thus the first step is legal.

For the second step we need $\forall x,y \in \Gamma,\Gamma'. x\sigma =
y\sigma \wedge x \neq y \Rightarrow \act_{\Lambda \cup \Lambda'}(x) =
\act_{\Lambda \cup \Lambda'}(y)$.  Note that since $(\Gamma \cup \Phi)
\cap (\Gamma' \cup \Phi')=\emptyset$ and $\sigma:\Gamma \rightarrow
\Gamma$, we have that $\sigma$ is the identity on $\Gamma'$, thus the
only $x,y$ such that $x\sigma = y\sigma \wedge x \neq y$ are in
$\Gamma$, thus the condition is satisfied since it was satisfied in
derivation \ref{eqn:merge-par}. Furthermore we have
$(\Gamma,\Gamma')\sigma=\Gamma\sigma,\Gamma'$ and
$(G_1|G'_1)\sigma=G_1\sigma|G'_1$.

We must now consider $\rho'$. We have:
$$\rho' = \mgu(\{(\n_{\Lambda \cup \Lambda'}(x))\sigma=
(\n_{\Lambda \cup \Lambda'}(y))\sigma|x\sigma= y\sigma \} \cup \{ x\sigma = y\sigma
| x(\pi \cup \pi')=y(\pi \cup \pi')\})$$

For what already said we have $\rho'= \mgu(\{(\n_{\Lambda}(x))\sigma=
(\n_{\Lambda}(y))\sigma|x\sigma= y\sigma \} \cup \{ x\sigma = y\sigma
| x\pi=y\pi\} \cup \{ x\sigma = y\sigma | x\pi'=y\pi'\})=\rho\pi'$
where $\rho$ is the one used in derivation \ref{eqn:merge-par}.

In particular, $\Lambda''(x)=\Lambda_1(x)$ if $x \in \Gamma\sigma$ and
$\Lambda''(x)=\Lambda'(x)\pi'=\Lambda'(x)$ since $\pi'$ is the identity
on representatives (see Definition \ref{defin:SHRtrans}). Thus
$\Lambda''=\Lambda_1 \cup \Lambda'$. Furthermore
$\pi''=\rho'|_{\Gamma\sigma,\Gamma'}=\pi_1 \cup \pi'$.

Also, $\Phi''=\Phi_1,\Phi'$ since it is determined by the first part
of transition.  Finally
$(G_2|G'_2)\sigma\rho'=G_2\sigma\rho'|G'_2\sigma\rho'=G_2\sigma\rho|G'_2\pi'=G_2\sigma\rho|G'_2$.
This proves that case.

We will now consider the composition of two (merge) rules, with
substitutions $\sigma$ and $\sigma'$ respectively.
Suppose we have a derivation of the form:

\begin{equation}\label{eqn:merge-merge}
\infer{\Gamma\sigma\sigma' \vdash G_1\sigma\sigma' \xrightarrow[]{\Lambda'', \pi''} \Phi'' \vdash G_2\sigma\rho\sigma'\rho'}
   {\infer{\Gamma\sigma \vdash G_1\sigma \xrightarrow[]{\Lambda', \pi'} \Phi' \vdash G_2\sigma\rho}
        {\Gamma \vdash G_1 \xrightarrow[]{\Lambda, \pi} \Phi \vdash G_2}
   }
\end{equation}

We want to be able to derive the same transition using just one inference step, with substitution $\sigma\sigma'$. We have:

\begin{equation}\label{eqn:merge}
\infer{\Gamma\sigma\sigma' \vdash G_1\sigma\sigma' \xrightarrow[]{\Lambda_1, \pi_1} \Phi_1 \vdash G_2\sigma\sigma'\rho_1}
   {\Gamma \vdash G_1 \xrightarrow[]{\Lambda, \pi} \Phi \vdash G_2}
\end{equation}

First of all we have to prove that the step is allowed. The required
condition is that $\forall x,y \in
\Gamma. x\sigma\sigma'=y\sigma\sigma' \land x \neq y \Rightarrow
\act_{\Lambda}(x)=\act_{\Lambda}(y)$. We have two cases. If
$x\sigma=y\sigma$ then the thesis follows from the analogous condition
of the first step of derivation \ref{eqn:merge-merge}. Otherwise we
can rewrite the condition as $\forall x,y \in
\Gamma. x\sigma\sigma'=y\sigma\sigma' \land x\sigma \neq y\sigma
\Rightarrow \act_{\Lambda}(x)=\act_{\Lambda}(y)$. Note that
$\act_{\Lambda'}(x\sigma)=\act_{\Lambda}(x)$ thus we can rewrite the
condition as $\forall x',y' \in \Gamma\sigma. x'\sigma'=y'\sigma'
\land x' \neq y' \Rightarrow \act_{\Lambda'}(x')=\act_{\Lambda'}(y')$
what is the condition for the second step of derivation
\ref{eqn:merge-merge}.

The main step now is to prove that
$\sigma\sigma'\rho_1=\sigma\rho\sigma'\rho'$.
We have:\\
$\rho=\mgu(\{(\n_{\Lambda}(x))\sigma=(\n_{\Lambda}(y))\sigma|x\sigma=y\sigma\} \cup \{x\sigma=y\sigma|x\pi=y\pi\})$\\
$\rho'=\mgu(\{(\n_{\Lambda'}(x))\sigma'=(\n_{\Lambda'}(y))\sigma'|x\sigma'=y\sigma'\} \cup \{x\sigma'=y\sigma'|x\pi'=y\pi'\})$\\
$\rho_1=\mgu(\{(\n_{\Lambda}(x))\sigma\sigma'=(\n_{\Lambda}(y))\sigma\sigma'|x\sigma\sigma'=y\sigma\sigma'\} \cup \{x\sigma\sigma'=y\sigma\sigma'|x\pi=y\pi\})$\\
In particular we have:
\begin{multline*}
\sigma\sigma'\rho_1\eqnum{1}\\
=\sigma\sigma'\mgu(\{(\n_{\Lambda}(x))\sigma\sigma'=(\n_{\Lambda}(y))\sigma\sigma'|x\sigma\sigma'=y\sigma\sigma' \land x \neq y\} \cup \\ \cup \{x\sigma\sigma'=y\sigma\sigma'|x\pi=y\pi\})\eqnum{2}\\
=\sigma\mgu(\{(\n_{\Lambda}(x))\sigma=(\n_{\Lambda}(y))\sigma|x\sigma\sigma'=y\sigma\sigma' \land x \neq y\} \cup\\ \cup \{x\sigma=y\sigma|x\pi=y\pi\} \cup \eqn(\sigma'))\eqnum{3}\\
=\sigma\mgu(\{(\n_{\Lambda}(x))\sigma=(\n_{\Lambda}(y))\sigma|x\sigma=y\sigma \land x \neq y\} \cup \\ \cup \{(\n_{\Lambda}(x))\sigma=(\n_{\Lambda}(y))\sigma|x\sigma\sigma'=y\sigma\sigma' \land x\sigma \neq y\sigma\} \cup \{x\sigma=y\sigma|x\pi=y\pi\} \cup \eqn(\sigma'))\eqnum{4}\\
=\sigma\mgu(\eqn(\rho) \cup \{(\n_{\Lambda}(x))\sigma=(\n_{\Lambda}(y))\sigma|x\sigma\sigma'=y\sigma\sigma' \land x\sigma \neq y\sigma\} \cup \eqn(\sigma'))\eqnum{5}\\
=\sigma\rho\mgu(\{(\n_{\Lambda}(x))\sigma\rho=(\n_{\Lambda}(y))\sigma\rho|x\sigma\sigma'=y\sigma\sigma' \land x\sigma \neq y\sigma\} \cup \eqn(\sigma')\rho)\eqnum{6}\\
=\sigma\rho\mgu(\{\n_{\Lambda'}(x\sigma)=\n_{\Lambda'}(y\sigma)|x\sigma\sigma'=y\sigma\sigma' \land x\sigma \neq y\sigma\} \cup \eqn(\sigma') \cup \eqn(\rho))\eqnum{7}\\
=\sigma\rho\sigma'\mgu(\{(\n_{\Lambda'}(x\sigma))\sigma'=(\n_{\Lambda'}(y\sigma))\sigma'|x\sigma\sigma'=y\sigma\sigma' \land x\sigma \neq y\sigma\} \cup \eqn(\rho)\sigma')\eqnum{8}\\
=\sigma\rho\sigma'\mgu(\{(\n_{\Lambda'}(x'))\sigma'=(\n_{\Lambda'}(y'))\sigma'|x'\sigma'=y'\sigma' \land x' \neq y'\} \cup\\ \cup \eqn(\pi')\sigma' \cup \eqn(\rho \setminus \pi')\sigma')\eqnum{9}\\
=\sigma\rho\sigma'\mgu(\{(\n_{\Lambda'}(x'))\sigma'=(\n_{\Lambda'}(y'))\sigma'|x'\sigma'=y'\sigma' \land x' \neq y'\} \cup\\ \cup \{x\sigma'=y\sigma'|x\pi'=y\pi'\} \cup \eqn(\rho \setminus \pi')\sigma')\eqnum{10}\\
=\sigma\rho\sigma'\rho'\mgu(\eqn(\rho \setminus \pi')\sigma'\rho')\eqnum{11}\\
=\sigma\rho\sigma'\rho'
\end{multline*}

We add some explanations for that (long) sequence of equations. Step
$1$ is just the definition of $\rho_1$. Step $2$ is allowed by Lemma
\ref{lemma:mgusubst}. Step $3$ is a simple mathematical
transformation. Step $4$ applies the definition of $\rho$. Step $5$ is
Lemma \ref{lemma:mgusubst} again. Step $6$ creates a new $\rho$ on the
outside using idempotence, then it brings it inside using Lemma
\ref{lemma:mgusubst} and uses idempotence again to delete it where it
is not necessary. It uses the definition of $\Lambda'$ too. Step $7$
is Lemma \ref{lemma:mgusubst} again. Steps $8$ and $9$ are trivial
mathematics. Step $10$ is another application of Lemma
\ref{lemma:mgusubst}. Finally, step $11$ is justified since the names
in the domain of $\rho \setminus \pi'$ are neither in the domain of
$\sigma'$ (since otherwise they would be in $\pi'$) nor in the domain
of $\rho'$ (since $\rho'$ is computed after having applied $\rho$,
which is idempotent). Thus an allowed $\mgu$ is a subset of $(\rho
\setminus \pi')\sigma'\rho'$ which can be deleted by idempotence.
\end{proof}

\begin{proof}[Proof of Theorem \ref{theorem:hoarecompleteness}]
Let us first consider the case of two productions for communication
actions. 

In order to apply them we need two sequential process edges
to be rewritten. Each production needs to be synchronized with at
least another one since each node is shared by exactly two
edges. Since process edges are connected only through amoeboids we can
have a synchronization only if the two actions done by process edges
are equal to two actions allowed by amoeboids. Thanks to Lemma
\ref{lemma:amoeboid}, they must be two complementary actions, \ie an
$in_n$ and an $out_n$.
Furthermore they have to be done on the same amoeboid, that
is on two names merged by the substitution $\sigma$ corresponding to
the amoeboids. 

Thus $P$ can be decomposed in the form $(\vec x)
P'\sigma$ where $P'=P'_1|P'_2|Q'$ with $P'_1$ and $P'_2$ sequential
processes which are translated into the rewritten edges.

We must
have $P'_1=\dots + u_1 \vec x'.P''_1$ and $P'_2=\overline{u_2} \vec
y'.P''_2+\dots$ (or swapped) with $|\vec x'|=|\vec y'|$. Furthermore
$\sigma$ merges $u_1$ and $u_2$ thus we have a transition $P
\rightarrow P'$ that corresponds to the synchronized execution of the
two prefixes.

The productions to be applied are thus forced except for the ones
inside the amoeboids, but the only difference among the choices (as
shown by Lemma \ref{lemma:amoeboid}) amounts to pseudoamoeboids and
exchanges between equivalent amoeboids in the result.  From Lemma
\ref{lemma:det} we know that the result of a transition is determined
up to injective renamings (and actions on isolated nodes) by the
starting graph and the productions chosen. Thus the transition that
corresponds to $P \rightarrow P'$ for Theorem
\ref{theorem:hoarecorrectness} is equal up to injective renamings to a
transition that differs from $\transl{P}_{\vec v ,\vec w}
\xrightarrow[]{\Lambda,id} \Gamma' \vdash G$ only for pseudoamoeboids
and substitutions of equivalent amoeboids.  The thesis follows.

The other case is analogous.
\end{proof}

\begin{lemma}\label{lemma:bigstep}
Let $A_1,\dots,A_n$ be a goal.

We want to build a big-step where the clause unified with $A_i$ is
$H_i \leftarrow B_i$, if any (some $A_i$ may not be replaced, in that
case as notation we use $B_i=A_i$). As a notational convention we use
$x_{i,1},\dots,x_{i,n_i}$ to denote the arguments of $A_i$ and
$a_{i,j}(x'_{i,j},\vec y'_{i,j})$ to denote the jth argument of $H_i$
if it is a complex term and $x'_{i,j}$ if it is a variable (note that
we have different names for the same variable, one for each
occurrence). All these are undefined if $A_i$ is not replaced,
$a_{i,j}$ and $\vec y'_{i,j}$ are undefined also if the jth argument
of $H_i$ is a variable.

Let $\theta_r$ be the mgu of the following set of
equations: 
$$\{ x'_{i,j}=x'_{p,q},\vec y'_{i,j}=\vec y'_{p,q} |
x_{i,j}=x_{p,q}\} \cup \{ x_{i,j}=x'_{i,j} | \textrm{the jth argument
of } H_i \textrm{ is } x'_{i,j}\}$$  

We will denote $x\theta_r$ with $[x]$. 

We will have a big-step of the form $A_1,\dots,A_n
\xRightarrow{\theta} G_1,\dots,G_n$ iff\\ 
$\forall i,p \in \{1,\dots,n\}. \forall j \in \{1,\dots,n_i\}. \forall q \in \{1,\dots,n_p\}.x_{i,j}=x_{p,q} \Rightarrow a_{i,j}=a_{p,q}$. 

Furthermore we have:

$\theta=\{a_{i,j}([x'_{i,j}],\vec {[y'_{i,j}]})/x_{i,j} | a_{i,j}
\textrm{ is defined}\} \cup \{ [v_k]/v_k | (v_k = x'_{i,j} \lor
v_k=\vec y'_{i,j}[l] \lor (v_k = x_{i,j} \land a_{i,j} \textrm{ is undefined
})) \land [v_k] \neq v_k \}$

$G_1,\dots,G_n=(B_1,\dots,B_n)\theta_r$

where $\vec{[y'_{i,j}]}[l]=[\vec{y'_{i,j}}[l]]$.

The big-step is determined (up to injective renamings) by
the choice of the clauses and of the atoms they are applied to.
\end{lemma}

\begin{proof}
We will prove a more general result by induction on the number of
``considered'' atoms, that is we consider an increasing chain of
derivations, and considered atoms are the ones that, if expanded in
the complete derivation, have already been expanded. For simplicity,
atoms are considered in numeric order, \ie at step $m$ atoms
$A_1,\dots,A_{m-1}$ have already been considered, and atom $A_m$
becomes considered.

We will prove that a computation of the form $A_1,\dots,A_n 
\xrightarrow[]{\theta} \mstar G_1,\dots,G_n$ where we substitute only atoms in 
the starting goal and where the atoms generated by considered atoms do not 
contain function symbols exists iff:\\
$\forall i,p \in \{1,\dots,m\}. \forall j \in \{1,\dots,n_i\}. \forall q \in \{1,\dots,n_p\}. x_{i,j}=x_{p,q} \Rightarrow a_{i,j}=a_{p,q}$ and that furthermore:
\begin{itemize}
\item $\theta=\{a_{i,j}([x'_{i,j}]',\vec {[y'_{i,j}]'})/x_{i,j} | a_{i,j} 
\textrm{ is defined} \land i \leq m\} \cup \{ [v_k]'/v_k | (v_k 
= x'_{i,j} \lor v_k=\vec y'_{i,j}[l] \lor (v_k = x_{i,j} \land a_{i,j} \textrm{
is undefined })) \land [v_k]' \neq v_k\}$;
\item $G_1,\dots,G_n=(B_1,\dots,B_m)\theta'_r,(A_{m+1},\dots,A_n)\theta$;
\end{itemize} 
where $\theta'_r$ is the $\mgu$ of the subset of $\eqn(\theta)$ containing only equalities 
between variables occurring in considered atoms and where $[x]'=x\theta'_r$.

Note that if all the atoms are considered we obtain the thesis.

Base case, $m=1$) We have the following transition:\\
$A_1,\dots,A_n \xrightarrow[]{\theta_1} (B_1,A_2,\dots,A_n) \theta_1$\\
where $\theta_1$ is an mgu of $\{A_1=H_1\}$. This transition exists iff 
$\theta_1$ exists.\\
We have:\\
$\theta_1=\mgu(\{ x_{1,j}=a_{1,j}(x'_{1,j},\vec y'_{1,j})|a_{1,j} \textrm{ 
is defined} \} \cup \{x_{1,j}=x'_{1,j}|a_{1,j} \textrm{ is 
undefined}\})$.\\
Note that if $x_{1,j}=x_{1,q} \nRightarrow a_{1,j}=a_{1,q}$ then the mgu 
does not exist (if $a_{i,j}$ and $a_{1,q}$ are both defined, otherwise 
they have to be both undefined since if just one of them is defined then a 
function symbol will remain in the considered part against the 
hypothesis).

If the condition is satisfied we have:
\begin{multline*}
\theta_1=\mgu(\{ x_{1,j}=a_{1,j}(x'_{1,j},\vec y'_{1,j})|a_{1,j} \textrm{ 
is defined} \} \cup \\ \cup \{ a_{1,q}(x'_{1,q},\vec y'_{1,q})= 
a_{1,j}(x'_{1,j},\vec y'_{1,j}) | x_{1,j}=x_{1,q}\} \cup 
\{x_{1,j}=x'_{1,j}|a_{1,j} \textrm{ is undefined}\})=\\
=\mgu(\{ x_{1,j}=a_{1,j}(x'_{1,j},\vec y'_{1,j})|a_{1,j} \textrm{ is 
defined} \} \cup\\ \cup \{ x'_{1,q}=x'_{1,j} , \vec y'_{1,q}=\vec y'_{1,j} 
| x_{1,j}=x_{1,q}\} \cup \{x_{1,j}=x'_{1,j}|a_{1,j} \textrm{ is 
undefined}\})
\end{multline*}
Note that the last part is exactly $\eqn(\theta'_r)$, thus $\theta'_r$ is its mgu.
 
Thus we have:\\
$\theta_1=\mgu(\{ x_{1,j}=a_{1,j}(x'_{1,j},\vec y'_{1,j})|a_{1,j} \textrm{ 
is defined} \} \cup \{ [v_k]'=v_k|[v_k]' \neq v_k\})$.\\
To have the real mgu we just need to apply $\theta'_r$ to the equations in 
the first part (note that $x_{1,j}$ is unified with some other variable 
only if $a_{1,j}$ is undefined thus the domain variables of the first part 
are not changed).\\ 
This proves the first part since this mgu exists and the second one since 
it has the wanted form.\\
The third part follows from the observation that 
$\theta_1|_{\n(B_1)}=\theta'_r|_{\n(B_1)}$.

Inductive case, $m \Rightarrow m+1$) Assume that $A_1,\dots,A_n
\xrightarrow[]{\theta_g} \mstar G_1,\dots,G_n$ is a logic computation where
we substitute only atoms in the starting goal, where atoms
$A_1,\dots,A_{m+1}$ are considered and where the atoms generated by
them do not contain function symbols. 

Let us take the subcomputation where only the first $m$ atoms have been
considered.

By inductive hypothesis we have that this part of the computation exists 
iff:\\
$\forall i,p \in \{1,\dots,m\}. \forall j \in \{1,\dots,n_i\}. \forall q \in \{1,\dots,n_p\}. x_{i,j}=x_{p,q} \Rightarrow a_{i,j}=a_{p,q}$\\
and that:
\begin{itemize}
\item $\theta=\{a_{i,j}([x'_{i,j}],\vec {[y'_{i,j}]})/x_{i,j} | a_{i,j} 
\textrm{ is defined } \land i \leq m \} \cup \{ [v_k]/v_k | (v_k = 
x'_{i,j} \lor v_k=\vec y'_{i,j}[l] \lor (v_k = x_{i,j} \land a_{i,j} \textrm{ is 
undefined })) \land [v_k] \neq v_k\}$;\\
\item $G_1,\dots,G_n=(B_1,\dots,B_m)\theta_r,(A_{m+1},\dots,A_n)\theta$.
\end{itemize}
We will now consider the atom $A_{m+1}$. If it is not substituted then
the thesis follows trivially (note that $A_{m+1}$ does not contain
variables substituted with a complex term by $\theta$, since this can
happen only if we have $x_{i,j}=x_{m+1,q}$ with $i<m+1$, $x_{i,j}$
defined and $x_{m+1,q}$ undefined, and this is forbidden; thus we have
$A_{m+1}\theta=A_{m+1}\theta_r$). Let us consider the case in which it
is substituted.

We will have a small step of the form:\\ 
$(G_1,\dots,G_m,A_{m+1},\dots,A_n)\theta \xrightarrow[]{\theta'} (\ 
G_1,\dots,G_{m+1},A_{m+2},\dots,A_n)\theta\theta'$\\
where $\theta'=\mgu(\{A_{m+1}\theta=H_{m+1}\})$ (note that we can assume 
that $\theta$ is also applied to $G_{m+1}$ since we can assume $\dom(\theta) \cap 
\n(G_{m+1})=\emptyset$).

We have:
\begin{multline*}
\theta'=\mgu(\{A_{m+1}\theta=H_{m+1}\})=\\
=\mgu(\{ x_{m+1,j}\theta = a_{m+1,j}(x'_{m+1,j},\vec y'_{m+1,j})|a_{m+1,j} 
\textrm{ is defined} \} \cup\\ \cup 
\{x_{m+1,j}\theta=x'_{m+1,j}|a_{m+1,j} \textrm{ is undefined}\})
\end{multline*}
For each binding we must consider two cases: either the variable
$x_{m+1,j}$ appears in already considered atoms (we call it an
\emph{old variable}) or it does not (we call it a \emph{new
variable}). In the second case $\theta$ is the identity on that
variable. In the first case if $a_{m+1,j}$ is defined then the mgu
exists iff we have $a_{m+1,j}=a_{p,q}$ where $a_{p,q}$ is the function
symbol in the binding for $x_{m+1,j}$ in $\theta$.  Note that if
$a_{p,q}$ is undefined then also $a_{m+1,j}$ must be undefined
otherwise the function symbol remains in the final goal.

Thus we will have:
\begin{multline*}
\theta'=\mgu(\{A_{m+1}\theta=H_{m+1}\})=\\ =\mgu(\{
a_{p,q}([x'_{p,q}],\vec {[y'_{p,q}]}) = a_{m+1,j}(x'_{m+1,j},\vec
y'_{m+1,j})|a_{m+1,j} \textrm{ is defined } \land \\ \land
x_{p,q}=x_{m+1,j} \land x_{m+1,j} \textrm{ is old} \} \cup
\{[x_{m+1,j}]=x'_{m+1,j}|a_{m+1,j} \textrm{ is undefined}\} \cup \\
\cup \{ x_{m+1,j}=a_{m+1,j}(x'_{m+1,j},\vec y'_{m+1,j})|a_{m+1,j}
\textrm{ is defined } \land x_{k+1,j} \textrm{ is new}\}=\\ =\mgu(\{
[x'_{p,q}]=x'_{m+1,j}, \vec{[y'_{p,q}]}=\vec y'_{m+1,j}|a_{m+1,j}
\textrm{ is defined } \land \\ \land x_{p,q}=x_{m+1,j} \land x_{m+1,j}
\textrm{ is old} \} \cup \{[x_{m+1,j}]=x'_{m+1,j}|a_{m+1,j} \textrm{
is undefined}\} \cup \\ \cup \{ x_{m+1,j}=a_{m+1,j}(x'_{m+1,j},\vec
y'_{m+1,j})|a_{m+1,j} \textrm{ is defined } \land x_{m+1,j}
\textrm{ is new}\}
\end{multline*}
where we use $[-]$ to denote the representative of the equivalence class 
according to $\theta$. We can reorder this substitution into:
\begin{multline*}
\mgu( \{ x_{m+1,j}=a_{m+1,j}(x'_{m+1,j},\vec y'_{m+1,j})|a_{m+1,j} 
\textrm{ is defined } \land x_{m+1,j} \textrm{ is new}\} \cup \\ \cup \{ [x'_{p,q}]=x'_{m+1,j}, 
\vec{[y'_{p,q}]}=\vec y'_{m+1,j}|a_{k+1,j} \textrm{ is defined } \land\\ \land
x_{p,q}=x_{m+1,j} \land x_{m+1,j} \textrm{ is old} \} \cup 
\{[x_{m+1,j}]=x'_{m+1,j}|a_{m+1,j} \textrm{ is undefined}\}
\end{multline*}
Note that the second part is a renaming that does not involve variables in 
the domain of the first part. Thus we can define new equivalence classes 
$[-]'$ according to this substitution $\theta'_r$ and put the substitution 
in the resolved form:
\begin{multline*}
\mgu( \{ x_{m+1,j}=a_{m+1,j}([x'_{m+1,j}]',[\vec y'_{m+1,j}]')|a_{m+1,j} 
\textrm{ is defined } \land x_{m+1,j} \textrm{ is new}\} \cup \\ \cup \{ [v_k]'=v_k|[v_k]' \neq v_k\})
\end{multline*}
Thus the mgu exists and the first part of the thesis is proved.

Let us consider the substitution $\theta_g=\theta\theta'$. Note that the 
renaming part of $\theta'$ substitutes variables according to the 
equivalence between variables in the $H_{m+1}$ and representatives of the 
corresponding variables in $A_{m+1}$ thus the composed substitution 
$\theta''_r$ maps each variable to the representative of the equivalence 
class that is defined by the union of the two sets of equations as 
required. Furthermore bindings with complex terms (the first part of the 
substitution) have disjoint domains and thus the union of them is made. Bindings coming 
from $\theta'$ have already the wanted representatives in the image, while 
to bindings in $\theta$ the renaming is applied, mapping variables into 
the representatives of their equivalence classes. Thus $\theta_g$ has the wanted form \wrt 
the equivalence classes determined by all the equivalences on variables.

We have:\\ 
$(G_1,\dots,G_{m+1},A_{m+2},\dots,A_n)\theta_g= 
(G_1,\dots,G_{m+1})\theta_r'',(A_{m+2},\dots,A_n)\theta_g$\\
as required 
since $(\dom(\theta_g) \setminus \dom(\theta''_r)) \cap 
\n(G_1,\dots,G_{m+1})=\emptyset$.

Note that this result does not depend on the order of application of 
clauses and that after having chosen which clauses to apply and to which 
atoms it is deterministic up to an injective renaming (which depends on 
the choice of names for new variables and on the choice of representatives 
for the equivalence classes).
\end{proof}

\begin{proof}[Proof of Theorem \ref{theorem:Tuostocorrectness}]
The proof is by rule induction.\\
Axioms)\\
Assume we have a production:\\
$x_1,\dots,x_n \vdash s(x_1,\dots,x_n) \xrightarrow[]{\Lambda,\pi} \Phi 
\vdash G$\\
where:\\
$\Lambda(x_i) = (a_i,\vec y_i) \ \forall i \in \{1 \dots n\}$\\
and $\pi:\{x_1,\dots,x_n\} \rightarrow \{x_1,\dots,x_n\}$ is an idempotent 
substitution. 

Then we have in $P$ a clause:\\
$s(a_1(x_i\pi,\vec y_1), \dots , a_n(x_n\pi,\vec y_n)) \leftarrow \transl{\Phi 
\vdash G}$\\
where we have $x_i\pi$ instead of $a_i(x_i\pi,\vec y_i)$ if 
$a_i=\epsilon$.
 
We can have an applicable variant of this clause for each injective 
renaming $\rho$ that maps each variable to a fresh one.
Let us consider the goal:\\
$\transl{ s(x_1, \dots, x_n) } = s(x_1, \dots, x_n)$\\
It unifies with the clause variant:\\
$s(a_1(x_i\pi,\vec y_1), \dots , a_n(x_n\pi,\vec y_n))\rho \leftarrow \transl{ 
\Phi \vdash G } \rho$\\
with mgu $\theta_{\rho}$:\\
$\theta_{\rho}=\{a_i(x_i\pi\rho,\vec y_i\rho)/x_i|a_i \neq \epsilon\} 
\cup \{x_i\pi\rho/x_i|a_i=\epsilon\}$.\\
Note that $\theta_{\rho}=\theta_{\rho}|_{\n(s(x_1,\dots,x_n))}$.
We can see that $\theta_{\rho}$ is associated to $x_1,\dots,x_n \vdash 
s(x_1,\dots,x_n) \xrightarrow[]{\Lambda,\pi} \Phi \vdash G$ as required.
Observe also that the result of the computation is:\\
$T = \transl{ \Phi \vdash G } \rho \theta_{\rho}=\transl{ \Phi \vdash G } 
\rho$\\
because $\n(G\rho) \cap \dom(\theta_{\rho})=\emptyset$.
This proves the thesis.
To deal with the missing possibilities for $\rho$ \wrt the 
theorem statement note that bindings in the last part of $\theta_{\rho}$ can also be 
resolved also as $x_i/x_i\pi\rho$. In that case we have no binding for 
$x_i$ in $\theta_{\rho}$. This is equivalent to defining $x_i\rho=x_i$, 
and this covers the missing cases for $\rho$.

Note finally that we used as clause the translation of the production and 
that we applied it to the translation of the rewritten edge.

Rule (par))
$$\frac{\Gamma \vdash G_1 \xrightarrow[]{\Lambda,
\pi} \Phi \vdash G_2 \quad \Gamma' \vdash G'_1
\xrightarrow[]{\Lambda',\pi'} \Phi' \vdash G'_2 \quad (\Gamma \cup
\Phi) \cap (\Gamma' \cup \Phi')=\emptyset}{\Gamma, \Gamma' \vdash
G_1|G'_1 \xrightarrow[]{\Lambda \cup \Lambda', \pi \cup \pi'} \Phi,
\Phi' \vdash G_2|G'_2}$$

By inductive hypothesis we have:
\begin{itemize}
\item $\transl{ \Gamma \vdash G_1 } \xRightarrow{\theta_{\rho}} T$\\
where $\theta_{\rho}$ is associated to $\Gamma \vdash G_1 
\xrightarrow[]{\Lambda,\pi} \Phi \vdash G_2$ and $T=\transl{ \Phi 
\vdash G_2 }\rho$;
\item $\transl{ \Gamma' \vdash G'_1 } \xRightarrow{\theta'_{\rho'}} T'$\\
where $\theta'_{\rho'}$ is associated to $\Gamma' \vdash G'_1 
\xrightarrow[]{\Lambda',\pi'} \Phi' \vdash G'_2$ and 
$T'=\transl{ \Phi' \vdash G'_2 }\rho'$.
\end{itemize}
In both cases by inductive hypothesis we used as clauses the translations 
of the productions used in the proof of the HSHR transition applied to the 
translations of the edges on which the productions were applied.\\

Thanks to Lemma \ref{lemma:injren} we can assume that the sets of 
variables used in the two computations are disjoint. Thanks to Lemma 
\ref{lemma:bigstep} we know that these big-steps exist iff 
$x_{i,j}=x_{p,q} \Rightarrow a_{i,j}=a_{p,q}$. Since the used variable 
sets are disjoint the same condition guarantees the existence of a 
big-step of the form:\\
$\transl{ \Gamma, \Gamma' \vdash G_1|G'_1 } 
\xrightarrow[]{\theta_{\rho}\theta'_{\rho'}} \mstar T,T'$\\
where we used as clauses the unions of the clauses used in the two smaller 
computations, applied to the same predicates.

We have that $\theta_{\rho}\theta'_{\rho'}$ is associated to:\\ 
$\Gamma, \Gamma' \vdash G_1|G'_1 \xrightarrow[]{\Lambda \cup \Lambda', \pi 
\cup \pi'} \Phi, \Phi' \vdash G_2|G'_2$.\\
We also have $\theta_{\rho}\theta'_{\rho'}=(\theta\theta')_{\rho\rho'}$ 
and thus:\\
$T,T'=\transl{\Phi \vdash G_2}\rho,\transl{\Phi' \vdash G'_2}\rho'=\transl{\Phi,\Phi' \vdash G_2|G'_2}\rho\rho'$
as required.

Rule (merge))
$$\frac{\Gamma \vdash G_1 \xrightarrow[]{\Lambda, 
\pi} \Phi \vdash G_2 \quad \forall x,y \in \Gamma. x\sigma = y\sigma \wedge x \neq y \Rightarrow \act_{\Lambda}(x) = \act_{\Lambda}(y)}{\Gamma\sigma \vdash G_1\sigma 
\xrightarrow[]{\Lambda', \pi'} \Phi' \vdash G_2\sigma\rho_g}$$
where $\sigma:\Gamma \rightarrow \Gamma$ is an idempotent substitution and:
\begin{enumerate}[(iii).]
\renewcommand{\theenumi}{(\roman{enumi})}
\item $\rho_g = \mgu(\{(\n_{\Lambda}(x))\sigma=
(\n_{\Lambda}(y))\sigma|x\sigma= y\sigma \} \cup \{ x\sigma = y\sigma
| x\pi=y\pi\})$ where $\rho_g$ maps names to representatives in $\Gamma \sigma$ whenever possible
\item $\forall z \in \Gamma. \Lambda'(z\sigma) = (\Lambda(z))\sigma\rho_g$
\item $\pi' = \rho_g|_{\Gamma\sigma}$
\end{enumerate}
where we used $\rho_g$ instead of $\rho$ to avoid confusion with the injective renaming $\rho$.
For inductive hypothesis we have:\\
$\transl{ \Gamma \vdash G_1 } \xRightarrow{\theta_{\rho}} T$\\
where $\theta_{\rho}$ is associated to $\Gamma \vdash G_1 
\xrightarrow[]{\Lambda,\pi} \Phi \vdash G_2$ and $T=\transl{ \Phi 
\vdash G_2 }\rho$ and we used as clauses the translations of the productions 
used in the proof of the HSHR transition applied to the translations of 
the edges on which the productions were applied.

Thanks to Lemma \ref{lemma:bigstep} we have that this computation exists 
iff:\\
$x_{i,j}=x_{p,q} \Rightarrow a_{i,j}=a_{p,q}$\\
and that:
\begin{itemize}
\item $\theta_r=\mgu(\{ x'_{i,j}=x'_{p,q},\vec y'_{i,j}=\vec y'_{p,q} | 
x_{i,j}=x_{p,q}\} \cup$\\ 
$\{ x_{i,j}=x'_{i,j} | \textrm{the jth argument of } H_i \textrm{ is } 
x'_{i,j}\})$\\
\item $\theta=\{a_{i,j}([x'_{i,j}],\vec {[y'_{i,j}]})/x_{i,j} | a_{i,j} 
\textrm{ is defined}\} \cup \{ [v_k]/v_k | (v_k = x'_{i,j} \lor 
v_k=\vec y'_{i,j}[l] \lor (v_k = x_{i,j} \land a_{i,j} \textrm{ is undefined })) 
\land [v_k] \neq v_k \}$\\
\item $T = (B_1,\dots,B_n)\theta_r$
\end{itemize}
where $\transl{ \Gamma \vdash G_1 }= A_1,\dots,A_n$ and we used the
naming conventions defined in Lemma \ref{lemma:bigstep}. In particular
$\rho$ maps each variable to its primed version.

We have:\\
$\transl{ \Gamma \vdash G_1 }\sigma=\transl{ \Gamma\sigma \vdash 
G_1\sigma }$

We want to find a big-step that uses the same clauses of the previous one 
applied to the same atoms, but starting from this new goal. We will use an 
overline to denote the components of the new big-step.
Thanks to Lemma \ref{lemma:bigstep} such a big-step exists iff:\\
$x_{i,j}\sigma=x_{p,q}\sigma \Rightarrow a_{i,j}=a_{p,q}$\\
but if $x_{i,j}=x_{p,q}$ this has already been proved, otherwise this is 
guaranteed by the applicability conditions of the rule.

We have to prove that $\overline{\theta}_{\overline{\rho}}$ is associated 
to:\\
$\Gamma\sigma \vdash G_1\sigma \xrightarrow[]{\Lambda', \pi'} \Phi' 
\vdash G_2\sigma\rho_g$.\\
From Lemma \ref{lemma:bigstep} we have:\\
$\overline{\theta}_{\overline{\rho}}=\{a_{i,j}([x'_{i,j}]',\vec 
{[y'_{i,j}]'})/x_{i,j}\sigma | a_{i,j} \textrm{ is defined}\} \cup \{ 
[v_k]'/v_k | (v_k = x'_{i,j} \lor v_k=\vec y'_{i,j}[l] \lor (v_k = x_{i,j} \land 
a_{i,j} \textrm{ is undefined })) \land [v_k]' \neq v_k \}$\\
where $[-]'$ maps each variable to the representative of the equivalence class 
given by:
\begin{multline*}
\{ x'_{i,j}=x'_{p,q},\vec y'_{i,j}=\vec y'_{p,q} | 
x_{i,j}\sigma=x_{p,q}\sigma\} \cup\\ \cup \{ x_{i,j}\sigma=x'_{i,j} 
| \textrm{the jth argument of } H_i \textrm{ is } x'_{i,j}\}
\end{multline*}
Note that $x_{i,j}\sigma$ is never unified with a primed variable unless 
$a_{i,j}$ is undefined.

Since $\Lambda'(x_{i,j}\sigma)=(\Lambda(x_{i,j}))\sigma\rho_g$ we need
to prove that
$\overline{\theta_r}=\theta_r\rho^{-1}\sigma\rho_g\overline{\rho}$,
that is that $[x']'=[y']'$ iff $x\sigma\rho_g=y\sigma\rho_g$.

From the hypothesis and from Lemma \ref{lemma:mgusubst} we have that
$\sigma\rho_g=\mgu(\{\n_{\Lambda}(x)= \n_{\Lambda}(y)|x\sigma=
y\sigma \} \cup \{ x = y | x\pi=y\pi\} \cup \eqn(\sigma))$. The first
and the third part are equal (adding primes) to the equations for
$[-]'$. As far as the second part is concerned, $\rho$ maps variables
merged by $\pi$ to the same primed variable, thus these equations
become trivial.

This proves that $\overline{\theta}_{\overline{\rho}}$ is associated to 
$\Gamma\sigma \vdash G_1\sigma \xrightarrow[]{\Lambda', \pi'} \Phi' 
\vdash G_2\sigma\rho_g$ as required.

By hypothesis and using Lemma \ref{lemma:bigstep} we have:\\
$\transl{ \Phi \vdash G_2 }\rho =(B_1,\dots,B_n)\theta_r$\\
The result of the new computation is:\\
$(B_1,\dots,B_n)\overline{\theta_r}$\\
Thanks to the properties of $\overline{\rho}$ we have:\\
$(B_1,\dots,B_n)\overline{\theta_r}=(B_1,\dots,B_n)\theta_r\rho^{-1}\sigma\rho_g\overline{\rho}=\transl{ 
\Phi \vdash G_2 }\rho\rho^{-1}\sigma\rho_g\overline{\rho}=\transl{ \Phi' \vdash G_2\sigma\rho_g }\overline{\rho}$.

Rule (idle))
$$\Gamma \vdash G \xrightarrow[]{\Lambda_{\epsilon},id} \Gamma \vdash G$$
The proof is analogous to the proof for axioms, using as clause 
corresponding to the production the translation of the (idle) rule.

Rule (new))
$$\frac{\Gamma \vdash G_1 \xrightarrow[]{\Lambda,\pi} 
\Phi \vdash G_2 \quad x \notin \Gamma \quad \vec y \cap (\Gamma \cup \Phi \cup \{x\})= \emptyset}{\Gamma,x \vdash G_1 \xrightarrow[]{\Lambda \cup 
\{(x,a,\vec y)\},\pi} \Phi' \vdash G_2}$$
For inductive hypothesis we have in $P$ the following big-step of 
Synchronized Logic Programming:\\
$\transl{ \Gamma \vdash G_1 } \xrightarrow[]{\theta_{\rho}} \mstar T$\\
where $\theta_{\rho}$ is associated to:\\ 
$\Gamma \vdash G_1 \xrightarrow[]{\Lambda,\pi} \Phi \vdash G_2$\\ 
and $T=\transl{ \Phi \vdash G_2 }\rho$
but because $x \notin \n(G_1)$ we have that $\theta_{\rho}$ is also 
associated to:\\ 
$\Gamma,x \vdash G_1 \xrightarrow[]{\Lambda \cup \{(x,a,\vec y)\},\pi} 
\Phi' \vdash G_2$\\
The thesis follows.
\end{proof}

\begin{proof}[Proof of Theorem \ref{theorem:Tuostocompleteness}]
The translation $\transl{ \Gamma \vdash G }$ can be written in the form 
$A_1,\dots,A_n$ where the $A_is$ are translations of single edges.
We want to associate a HSHR production to each $A_i$.
For edges associated to atoms that are rewritten in $\transl{ \Gamma 
\vdash G } \xRightarrow{\theta} T$ we use an instance of the axiom that 
corresponds to the clause used to rewrite it, otherwise the rule obtained 
from rule (idle) applied to that edge and to nodes that are all distinct.

Let $T_1,\dots,T_n$ be the heads of these rules. We choose the names in 
the following way: for each first occurrence of a variable in $A_1,\dots,A_n$ 
we use the same name for the node in the corresponding position, we use 
new names for all other occurrences. Note that there exists a substitution 
$\sigma$ such that $(T_1,\dots,T_n)\sigma = A_1,\dots,A_n$ and that 
$\sigma$ is idempotent. Note also that for each $i$ all names in $T_i$ are 
distinct.

For each $i$ we can choose an instance $R_i$ of the associated rule with 
head $T_i$ such that for each $i,j, i \neq j$ we have $\n(R_i) \cap 
\n(R_j) = \emptyset$.
As notation we use:\\
$R_i=\Gamma_i \vdash G_i \xrightarrow[]{\Lambda_i, \pi_i} \Phi_i \vdash 
G'_i$\\
Since all the rules have a disjoint set of names we can apply $n-1$ times 
rule (par) in order to have:\\
$\bigcup_i \Gamma_i \vdash \bigpar_i G_i \xrightarrow[]{\bigcup_i \Lambda_i, 
\bigcup_i \pi_i} \bigcup_i \Phi_i \vdash \bigpar_i G'_i$\\
Now we want to apply rule (merge) with substitution $\sigma$. We can do it 
since $\sigma$ is idempotent.
We have to verify that $x\sigma=y\sigma \land x \neq y \Rightarrow 
(\bigcup_i \Lambda_i)(x)=(a, \vec v) \land (\bigcup_i \Lambda_i)(x)=(a, 
\vec w)$. This happens thanks to Lemma \ref{lemma:bigstep}.
Thus we obtain a rule of the form:\\ 
$\n(A_1,\dots,A_n) \vdash A_1,\dots,A_n \xrightarrow[]{\Lambda, \pi} \Phi 
\vdash G'$\\
Thanks to Theorem \ref{theorem:Tuostocorrectness} we can have in $P$ the 
following big-step of Synchronized Logic Programming:\\
$\transl{ \n(A_1,\dots,A_n) \vdash A_1,\dots,A_n } 
\xRightarrow{\theta'_{\rho}} T'$\\
for every $\rho$ that satisfies the freshness conditions. Furthermore 
$\theta'_{\rho}$ is associated to $\n(A_1,\dots,A_n) \vdash A_1,\dots,A_n 
\xrightarrow[]{\Lambda,\pi} \Phi \vdash G'$ and $T'=\transl{ \Phi 
\vdash G' }\rho$. Finally we used as clauses the translations of the 
productions used in the proof of the HSHR rewriting.
Note that $\transl{ \n(A_1,\dots,A_n) \vdash A_1,\dots,A_n }= \transl{ 
\Gamma \vdash G}$. Since the result of a big-step is determined up to 
injective renaming by the starting goal and the used clauses (see Lemma 
\ref{lemma:bigstep}) then we must have $\theta=\theta'_{\rho}\rho'$ and 
$T=T'\rho'$ for some injective renaming $\rho'$. Note that 
$\rho\rho'$ satisfies the freshness conditions (since variables 
generated in logic programming are always fresh) and we also have 
$\theta=\theta''_{\rho\rho'}$.\\
Thus we have that $\theta$ is associated to $\Gamma \vdash G 
\xrightarrow[]{\Lambda,\pi} \Phi \vdash G'$ and that 
$T=T'\rho=\transl{\Phi \vdash G'}\rho\rho'$.
This proves the thesis.
\end{proof}
\end{document}